\documentclass[review,12pt,sort&compress]{elsarticle}
\usepackage[utf8]{inputenc}
\usepackage{siunitx}
\usepackage{geometry}
\usepackage{comment}
\usepackage{epsfig}
\makeatletter \def\ps@pprintTitle{  \let\@oddhead\@empty  \let\@evenhead\@empty  \def\@oddfoot{\small{\today\hfill\thepage}}  
\def\@evenfoot{\thepage\hfill}} 
\makeatother
\usepackage{amsfonts,amssymb,amsmath,amsthm,amsbsy}
\usepackage{graphicx}
\usepackage[T1]{fontenc}
\usepackage[utf8]{inputenc}
\usepackage{listings}
\usepackage{appendix}
\usepackage{minted}
\usepackage{lineno}
\usepackage{color}
\definecolor{mygreen}{RGB}{28,172,0}
\definecolor{mylilas}{RGB}{170,55,241}
\usepackage{mathtools}
\numberwithin{equation}{section}
\usepackage{caption}
\usepackage{subcaption}
\usepackage{subfiles}
\usepackage[colorlinks=true,citecolor=blue,urlcolor=blue]{hyperref}
\usepackage{silence}
\usepackage{amsmath}
\numberwithin{equation}{section}
\usepackage{pgfplots}
\usepackage{tikz}
\pgfplotsset{compat=1.18}

\begin{document}

\begin{frontmatter}
\title{Linearized instability of Couette flow in stress-power law fluids\tnoteref{t1}}
\tnotetext[t1]{This work is dedicated to the memory of Prof. \ K.\ R.\ Rajagopal, whose guidance continues to inspire us.}
\author[1]{Krishna Kaushik Yanamundra}
\author[2]{Lorenzo Fusi}
\address[1]{Department of Mechanical Engineering, Texas A\&M University, College Station, TX 77843, USA}
\address[2]{Dipartimento di Matematica e Informatica ``U. Dini'', Universit\`{a}  degli Studi di Firenze, 50134, Firenze, Italy}

\begin{abstract}
This paper examines the linearized stability of plane Couette flow for stress-power law fluids, which exhibit non-monotonic stress–strain rate behavior. The constitutive model is derived from a thermodynamic framework using a non-convex rate of dissipation potential. Under velocity boundary conditions, the system may admit three steady-state solutions. Linearized stability analysis reveals that the two solutions on ascending constitutive branches are unconditionally stable, while the solution on the descending branch is unconditionally unstable. 
For mixed traction-velocity boundary conditions, the base state is unique. Stability depends solely on whether the prescribed traction lies on an ascending (stable) or descending (unstable) branch of the constitutive curve. The results demonstrate that flow stability in these complex fluids is fundamentally governed by both boundary conditions and constitutive non-monotonicity.
\end{abstract}

\begin{keyword}
\texttt{Stress power law fluid \sep implicit constitutive relations \sep non-monotonic stress strain rate response \sep plane Couette flow \sep linear stability analysis \sep Orr Sommerfeld eigenvalue problem \sep pseudospectral collocation method \sep shear banding \sep discontinuous shear thickening}
\end{keyword}
\end{frontmatter}

\section{Introduction}
The class of Stokesian fluids is defined as those fluids whose mechanical response is modeled by expressing the Cauchy stress tensor $\mathbf{T}$ as a function of the symmetric part of the velocity gradient $\mathbf{D}$, i.e., $\mathbf{T} = \mathbf{f}(\mathbf{D})$ (see \cite{Stokes1845TCPS,Truesdell1950JMPA}). While such a prescription is broad, it cannot adequately model fluids whose steady-state stress-strain rate response is non-monotonic, wherein a given strain rate may correspond to multiple stress values (see \cite{Boltenhagen1997PRL,Macias2001JNNFM,Bertrand2002PRE,Lopez2010JCIS,Wilkins2006EPJE}). Such fluids also exhibit markedly different responses in stress controlled and strain rate controlled experiments. In those cases, stress can no longer be expressed as a function of the symmetric part of the velocity gradient. Moreover, since the stress (traction) is the cause and motion (kinematics) the effect, prescribing the stress as a function of kinematical quantities is contrary to causality.\footnote{For a detailed discussion on the role of causality in the development of constitutive relations, see \cite{Rajagopal2003AM,Rajagopal2006JFM}.}

In view of this, Rajagopal \cite{Rajagopal2003AM,Rajagopal2006JFM} generalized the class of constitutive relations to allow for an implicit relationship between the Cauchy stress and the symmetric part of the velocity gradient, i.e. $\mathbf{g}(\mathbf{T}, \mathbf{D}) = \mathbf{O}$. While this broader definition contains the class of Stokesian fluids as a special case, it also naturally accommodates models that cannot be expressed within that class, such as fluids with pressure-dependent viscosity \cite{Hron2001PRSA}, and the widely used Bingham \cite{Oldroyd1947MPCPS} and Herschel–Bulkley \cite{Herschel1926KZ} models. A particularly interesting subset of these implicit constitutive relations, which may be viewed as a natural complement to the classical prescription, is the class of fluids whose symmetric part of the velocity gradient $\mathbf{D}$ is given explicitly in terms of Cauchy stress, $\mathbf{D} = \mathbf{h}(\mathbf{T})$. Clearly $\mathbf{f}$ is the inverse function of $\mathbf{h}$ whenever an inverse exists, but the utility of this formulation lies precisely in its ability to model fluids for which $\mathbf{h}$ is not bijective. Moreover, such a prescription is in keeping with the notion of causality.

Malek et al. \cite{Malek2010IJES} introduced one such constitutive relation wherein the symmetric part of the velocity gradient $\mathbf{D}$ was prescribed as a ``power-law'' of the deviatoric stress. For certain parameter values, this relation becomes non-monotonic and hence non-invertible. They observed that in Couette flows velocity boundary conditions may yield one, two, or no solutions, whereas prescribing a traction boundary condition always yields a unique solution. A similar dependence on boundary conditions was reported by Srinivasan and Karra \cite{Srinivasan2015IJNLM} for flows between eccentrically rotating disks.

Le Roux and Rajagopal \cite{LeRoux2013AppMath} generalized this model to capture the S-shaped, non-monotonic steady-state stress–strain rate response observed in several experiments \cite{Boltenhagen1997PRL,Macias2001JNNFM,Bertrand2002PRE,Lopez2010JCIS,Wilkins2006EPJE}, a behaviour commonly associated with discontinuous shear thickening and shear banding in complex fluids (see \cite{Olmsted2008RA,Morris2020ARFM}). This model has since been examined in a number of flow configurations, including squeeze flow \cite{Fusi2021F}, flows between eccentrically rotating disks \cite{Fusi2023IJNM,Yanamundra2024POF}, and pressure driven flows \cite{Almqvist2024AppMath,GomezConstante2019PRSA}.

Blechta et al. \cite{Blechta2020SIAM} and Rajagopal \cite{Rajagopal2023CR} have proposed other variants of stress power law type models which show stress limiting response. Yanamundra et al. \cite{Yanamundra2024POF} studied a generalization that encompasses many of these variants, and in the present work we derive such a generalized stress power-law model within the thermodynamic framework of Rajagopal and Srinivasa \cite{Rajagopal2000JNNFM,Rajagopal2008ZAMP} by constructing a rate of dissipation potential that is non-convex for a range of parameters.

The fact that these constitutive relations may admit non-unique solutions renders a stability analysis indispensable in identifying which among these states are physically admissible, and under what conditions. While some initial progress has been made in this direction, with studies examining the stability of the quiescent state and of flows past a porous plate \cite{Rajagopal2021AES,Fusi2022EJMB}, much remains to be understood about the stability of non-trivial base flows. 

Since many of the experiments in which the non-monotonic response was observed were conducted in cylindrical Couette geometries, it is natural to consider the stability of Taylor–Couette flow. However, Taylor–Couette flow is susceptible to a wide range of instabilities even for linearly viscous fluids \cite{Taylor1923PTRSA}, whereas plane Couette flow is unconditionally stable \cite{Romanov1973FAA}. Therefore, to distinguish instabilities induced by geometry from those arising from the constitutive prescription, we consider the stability of plane Couette flow.

The organization of the paper is as follows. In \S2, we present the kinematic and thermodynamic preliminaries and derive the generalized stress power–law model by maximizing the rate of entropy production subject to the second law of thermodynamics as a constraint. In \S3, we introduce an appropriate scaling and non–dimensionalize the governing equations. In \S4, we analyze planar Couette flow and obtain exact base–flow solutions for both velocity and traction boundary conditions. In \S5, we then linearize the governing equations about these base states, and arrive at an Orr–Sommerfeld–type eigenvalue problem. The pseudospectral collocation method used for its numerical solution and the corresponding stability results are discussed in \S6 and \S7, respectively. Finally, in \S8, we summarize the key conclusions and outline potential directions for future work.

\section{Constitutive Relations}
Let $\kappa_R(\mathcal{B})$ denote a stress-free reference configuration of the body $\mathcal{B}$, and let $\kappa_t(\mathcal{B})$ denote its configuration at current time $t$. The motion of the body from $\kappa_R(\mathcal{B})$ to $\kappa_t(\mathcal{B})$ is assumed to be a diffeomorphic mapping given by $\mathbf{\chi}(\mathbf{X},t)$, which assigns each material point $\mathbf{X}$ in the reference configuration a position $\mathbf{x}$ in the current configuration. The deformation gradient is defined as 
\begin{equation}
\mathbf{F} := \frac{\partial \mathbf{\chi}}{\partial \mathbf{X}},    
\end{equation}
and the velocity of the body as 
\begin{equation}
\mathbf{v} := \frac{\partial \mathbf{\chi}}{\partial t}.    
\end{equation}
The velocity gradient $\mathbf{L}$ and its symmetric part $\mathbf{D}$ are then given by
\begin{align}
\mathbf{L} := \frac{\partial \mathbf{v}}{\partial \mathbf{x}} = \dot{\mathbf{F}}\mathbf{F}^\mathrm{-1},& &\text{and}& &\mathbf{D} := \frac{1}{2}(\mathbf{L} + \mathbf{L}^\mathrm{T}).    
\end{align}
Let $\rho$ denote the mass density, $\epsilon$ the specific internal energy, $\eta$ the specific entropy, $\theta$ the temperature, $\mathbf{T}$ the Cauchy stress, and $\mathbf{q}$ the heat flux in the current configuration $\kappa_t$. The local form of the second law of thermodynamics may be expressed as
\begin{align}
\mathbf{T}\cdot\mathbf{D} - \rho\dot{\epsilon} + \rho\theta\dot{\eta} - \frac{1}{\theta}\mathbf{q}.\mathrm{grad}\,\theta = \zeta \geq 0, \label{Eqn2.2}
\end{align}
where, $\zeta$ signifies the rate of entropy production per unit volume, and the dot denotes the inner product of tensors, i.e. $\mathbf{M}\cdot\mathbf{N} := \mathrm{tr}(\mathbf{M}\mathbf{N}^\mathrm{T})$, $\mathrm{tr}(.)$ being the trace operator. Specific Helmholtz free energy $\psi$ is related to the specific internal energy $\epsilon$ through a Legendre transform given by
\begin{align}
\psi = \epsilon - \theta\eta.
\end{align}
Using this relation, \eqref{Eqn2.2} may be rewritten in terms of $\psi$ as
\begin{align}
\mathbf{T}\cdot\mathbf{D} - \rho\dot{\psi} - \rho\eta\dot{\theta}  - \frac{1}{\theta}\mathbf{q}.\mathrm{grad}\,\theta = \zeta \geq 0. \label{Eqn2.6}
\end{align}
Under the idealization that the fluid under consideration is purely viscous and therefore incapable of storing energy, the material time derivative of the free energy vanishes, i.e.\ $\dot{\psi}=0$. Furthermore, if the process is assumed to be both isothermal and homothermal, the inequality \eqref{Eqn2.6} simplifies to 
\begin{align}
\mathbf{T}\cdot\mathbf{D} = \zeta \geq 0.\label{Eqn2.7}
\end{align}
If we assume that the body has a uniform mass density $\rho$ and assume that the body can only undergo isochoric deformations, then from balance of mass, 
\begin{align}
\mathrm{div}(\mathbf{v}) = \mathrm{tr}(\mathbf{D}) = 0.
\end{align}
We decompose $\mathbf{T}$ into its spherical and deviatoric parts as 
 \begin{align}
 \mathbf{T}= \frac{1}{3}\mathrm{tr}(\mathbf{T)}\mathbf{I} + \mathbf{S},\label{Eqn0}
 \end{align}
 and since $\mathbf{D}$ is deviatoric, the inequality \eqref{Eqn2.7} further reduces to 
\begin{align}
\mathbf{S}\cdot\mathbf{D} = \zeta \geq 0.\label{Eqn2.7a}
\end{align}
Following Rajagopal and Srinivasa \cite{Rajagopal2000JNNFM}, we derive the constitutive relations by prescribing a specific form for the rate of dissipation, $\zeta$, which in the present case is taken to be a function of the deviatoric part of the Cauchy stress, $\hat{\zeta}(\mathbf{S})$, and by maximizing it subject to the constraints (\ref{Eqn2.7a}) and $\mathrm{tr}(\mathbf{S}) = 0$. Hence, we introduce the augmented function
\begin{align}
\Phi = \hat\zeta + \lambda_1(\hat\zeta - \mathbf{S}\cdot\mathbf{D}) + \lambda_2(\mathbf{I}\cdot\mathbf{S}),
\end{align}
where $\lambda_1$ and $\lambda_2$ are the Lagrange multipliers. Differentiating $\Phi$ with respect to $\mathbf{S}$ we obtain
\begin{align}
\frac{\partial \Phi}{\partial \mathbf{S}}
= (1+\lambda_1)\,\frac{\partial \hat{\zeta}}{\partial \mathbf{S}}
-\lambda_1\,\mathbf{D}
+\lambda_2\,\mathbf{I}
= \mathbf{0}.
\end{align}
Hence
\begin{align}
\mathbf{D}
= \frac{1+\lambda_1}{\lambda_1}\,\frac{\partial \hat{\zeta}}{\partial \mathbf{S}}
+\frac{\lambda_2}{\lambda_1}\,\mathbf{I}. \label{Eqn2.13}
\end{align}
Taking inner product of (\ref{Eqn2.13}) with $\mathbf{S}$ gives
\begin{align}
\frac{1 + \lambda_1}{\lambda_1} = \frac{\hat\zeta}{\frac{\partial \hat{\zeta}}{\partial \mathbf{S}}\cdot\mathbf{S}}\;,
\end{align}
and since $\mathrm{tr}(\mathbf{D})=0$, 
\begin{align}
\frac{\lambda_2}{\lambda_1} = -\frac{1 + \lambda_1}{3\lambda_1}\;\mathrm{tr}\left(\frac{\partial \hat{\zeta}}{\partial \mathbf{S}}\right).
\end{align}
Let us consider the rate of dissipation potential of the form 
\begin{align}
\hat\zeta(\mathbf{S}) = \alpha\left(a + \left(b + \beta\; \mathrm{tr}(\mathbf{S}^2)\right)^n\right)\mathrm{tr}(\mathbf{S}^2), \label{Eqn2.16}
\end{align}
where $\alpha \in \mathbb{R}^+$, $a \in \mathbb{R}^+_0$, $\beta \in \mathbb{R}^+_0$, $b \in \{0,1\}$, and 
\begin{align}
      n \in \begin{cases}
        \left[-\frac{1}{2},\infty\right), &\text{$b=0$}
        \\
        \left(-\infty,\infty\right), &\text{$b=1$}
        \end{cases}
\end{align}
such that $\hat\zeta$ is always non-negative, and it is 0 only when $\mathbf{S} = \mathbf{O}$, i.e. the zero tensor. The functional form (\ref{Eqn2.16}) depends solely on the second principal invariant of $\mathbf{S}$, thereby ensuring material isotropy and invariance under Galilean transformations. The potential chosen here is noteworthy in that it is non-convex for certain material parameters. In much of the literature, for instance  \cite{Rajagopal2000JNNFM,Karra2009AM,Karra2010AP,Narayan2021PST}, the dissipation potential is assumed to be strictly convex, thereby ensuring the uniqueness of the maximizer of the rate of dissipation potential, whereas in this case, when the potential is non-convex, the extremization process yields non-unique maximizers. This departure from standard assumptions allows us to model the non-monotonic response where multiple values of stress can cause the same strain rate.\\
Now, from (\ref{Eqn2.13}) and (\ref{Eqn2.16}) we get
\begin{subequations}
\begin{align}
    \mathbf{T} &= -p\mathbf{I}+\mathbf{S},\\
    \mathbf{D} &= \alpha \left(a + \left(b + \beta\;\mathrm{tr}(\mathbf{S}^2)\right)^n\right)\mathbf{S}, \label{Eqn1}
\end{align}
\end{subequations}
where $p$ is the mechanical pressure defined as the negative of the mean normal stress.

Let us qualitatively analyze the model for different parametric values. For $b = 1$ as $n$ varies the generalized stress power-law model behaves as follows:
\begin{align}
      n \in \begin{cases}
        \left[-\frac{1}{2},0\right): &\text{(\ref{Eqn1}) exhibits stress-thickening}\\
        \{0\}: &\text{(\ref{Eqn1}) reduces to a linearly viscous model}\\
        \left(0,\infty\right): &\text{(\ref{Eqn1}) exhibits stress-thinning}
        \end{cases}
\end{align}

When $b = 1$ and $n<-\frac{1}{2}$, nature of (\ref{Eqn1}) depends on $a$ as follows (Refer to Le Roux and Rajagopal \cite{LeRoux2013AppMath} for proof):
\begin{align}
      a \in \begin{cases}
        \{0\}: &\text{non-monotonic with one inflection point}\\
        \left(0,2\left(\frac{\left|2n+1\right|}{2\left(1-n\right)}\right)^{\left(1-n\right)}\right): &\text{S-type non-monotonicity with two inflection points}\\
        \left[2\left(\frac{\left|2n+1\right|}{2\left(1-n\right)}\right)^{\left(1-n\right)},\infty\right): &\text{monotonically increasing}
        \end{cases}
\end{align}
For $b = 0$ and $n<0$, the generalized fluidity, $\alpha_g(\|\mathbf{S}\|)$ given by $\alpha \left(a + \left(b + \beta \|\mathbf{S}\|^2\right)^n\right)$ in (\ref{Eqn1}) tends to $\infty$ as $\|\mathbf{S}\| \rightarrow 0$ behaving like an Euler fluid and asymptotically reaches $a\alpha$ as $\|\mathbf{S}\| \rightarrow \infty$, akin to a linear model. When $b=0$, and $n \rightarrow \infty$, the model shows ``stress-limiting'' behaviour. In this paper, we limit our analysis to the case where $\mathbf{D}$ varies non-monotonically with $\mathbf{S}$.

\captionsetup[subfigure]{labelformat=simple, labelsep=space}
\renewcommand{\thesubfigure}{(\roman{subfigure})}
\begin{figure}[t]
    \centering
    \includegraphics[width=\linewidth]{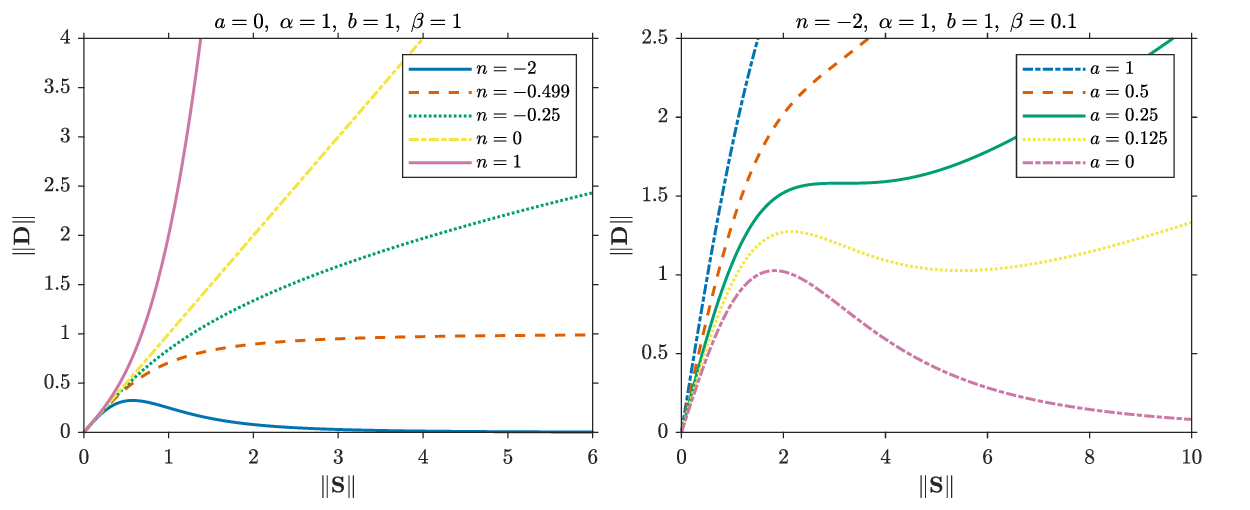}
    \caption{
    Constitutive response of the generalized stress power--law model.
    \textit{Left:} $\|\mathbf{D}\|$ as a function of $\|\mathbf{S}\|$ for representative values of $n$ with $a=0$, $\alpha=1$, $b=1$, $\beta=1$.
    \textit{Right:} Variation with $a$ for $n=-2$, $\alpha=1$, $b=1$, $\beta=0.1$, illustrating the emergence of non-monotonicity and the regime in which multiple solutions may arise.
    }
    \label{fig:constitutive}
\end{figure}

We solve the constitutive relation given by (\ref{Eqn1}) along with balance of linear momentum (neglecting body forces) given by 
\begin{align}
\rho \frac{d\mathbf{v}}{dt} = -\text{grad}(p) + \text{div}(\mathbf{S})\label{Eqn3}
\end{align}
where, $\frac{d}{dt} (.)$ is the material time derivative.

\section{Non-dimensionalization}
\noindent To non-dimensionalize the governing equations, we introduce the following scaled quantities:
\begin{align*}
&\overline{\mathbf{v}} = \frac{1}{v^*}\mathbf{v},&&\overline{\mathbf{x}} = \frac{1}{x^*}\mathbf{x},& &\overline{t} = \frac{v^*}{x^*}t,\\
&\overline{\mathbf{S}} = \frac{\alpha x^*}{v^*}\mathbf{S},& &\overline{\mathbf{T}} = \frac{\alpha x^*}{v^*}\mathbf{T},& &\overline{p} = \frac{\alpha x^*}{v^*}p,\\
&\overline{\mathbf{D}} = \frac{x^*}{v^*}\mathbf{D},&&\mathrm{Re} = \rho v^*x^*\alpha,& &\overline{\beta} = \left(\frac{v^*}{\alpha x^*}\right)^2\beta.
\end{align*}
Here, $x^*$ and $v^*$ denote characteristic length and velocity scales, respectively, $\rho$ is the density of the fluid and $\alpha$ is the fluidity parameter. The non-dimensional quantity $\rho\alpha v^*x^*$ can be defined as the Reynolds number (Re). The non-dimensional term $\overline{\beta}$ can then be expressed as $\Gamma \mathrm{Re}^2$, where 
\begin{align}
    \Gamma = \frac{\beta}{\rho^2\alpha^4{x^*}^4}
\end{align} 
is also non-dimensional and depends only on the material and geometric parameters.
Governing equations in the non-dimensional form then become (dropping the overbar)
\begin{subequations}
\begin{align} 
&\mathrm{Re}\frac{d\mathbf{v}}{dt} = -\text{grad}(p) + \text{div}(\mathbf{S}),\\
&\mathbf{D} = \left(a + \left(b + \Gamma \mathrm{Re}^2 \|\mathbf{S}\|^2\right)^n\right)\mathbf{S}.
\end{align}
\end{subequations}

\section{Planar-Couette base flow}
Let us consider the flow between two parallel planes separated by a distance $2h$. We use a right-handed Cartesian frame with its origin mid-way between the planes, with the $y$ coordinate normal to the planes and $x$ coordinate along the planes. We then choose the characteristic length $x^*$ as $h$ so that domain maps to the interval $[-1,1]$, which is convenient while using spectral methods to solve our resulting system of equations. Assuming that the base flow is planar, let us consider the following ansatz for the non-dimensional velocity $\mathbf{v}^{(b)}$ and the non-dimensional extra stress $\mathbf{S}^{(b)}$ as
\begin{align}
    \mathbf{v}^{(b)} = \begin{bmatrix}
        v^{(b)}_x(y)\\
        0
    \end{bmatrix},& &\text{and}& &\mathbf{S}^{(b)} = \begin{bmatrix}
        0 & s^{(b)}_{xy}(y)\\
        s^{(b)}_{xy}(y) & 0
    \end{bmatrix}.
\end{align}
It follows from the above ansatz that the velocity gradient of the base flow $\mathbf{L}^{(b)}$ can be written as  
\begin{align}
    \mathbf{L}^{(b)} &= \begin{bmatrix}
        0 & v^{(b)}_{x,y}\\
        0 & 0
    \end{bmatrix},
\end{align}
where $(.)_{,y}$ denotes derivative with respect to $y$. $\mathbf{L}^{(b)}$ can be decomposed into its symmetric and skew-symmetric parts $\mathbf{D}^{(b)}$ and $\mathbf{W}^{(b)}$ respectively, such that
\begin{align}
    \mathbf{D}^{(b)} = \frac{1}{2}\begin{bmatrix}
        0 & v^{(b)}_{x,y}\\
        v^{(b)}_{x,y} & 0
    \end{bmatrix},& &\text{and}& &\mathbf{W}^{(b)} = \frac{1}{2}\begin{bmatrix}
        0 & v^{(b)}_{x,y}\\
        -v^{(b)}_{x,y} & 0
    \end{bmatrix}.
\end{align}
Assuming there is no pressure gradient, the simplified governing equations for the base flow then become - 
\begin{align}
    &2\left(a + \left(b + 2\Gamma \mathrm{Re}^2{s^{(b)}_{xy}}^2\right)^n\right)s^{(b)}_{xy} = v^{(b)}_{x,y}\label{4.4}\\
    &s^{(b)}_{xy,y} = 0
\end{align}
It is straightforward to obtain an exact general solution for the above system as
\begin{align}
    s^{(b)}_{xy}(y) &= c_1, \label{Eqn3.6}\\
    v^{(b)}_{x}(y) &= 2\left(a + \left(b + 2\Gamma \mathrm{Re}^2c_1^2\right)^n\right)c_1y + c_2. \label{Eqn3.7}
\end{align}
Where $c_1$ and $c_2$ are constants of integration, which are to be determined from the specific boundary conditions.
\subsection{Boundary Conditions}
\subsubsection{Case 1: Velocity Boundary Conditions}\label{BCV}
Let us consider the case where the upper plane slides with a constant velocity $v_u$, and the lower plate with a velocity $v_l$. We shall use these quantities to define a Reynolds number corresponding to the upper and lower planes as 
\begin{align}
    &\mathrm{Re}_u = \rho\alpha h v_u, &\text{and}& &\mathrm{Re}_l = \rho\alpha h v_l.
\end{align}
If we then define our characteristic velocity scale $v^*$ as $(v_u - v_l)$, we could express our Reynolds number as a difference of the Reynolds numbers corresponding to the top and bottom planes 
\begin{align}
    &\mathrm{Re} = \rho\alpha h (v_u -v_l) = \mathrm{Re}_u - \mathrm{Re}_l.
\end{align}
The non-dimensional boundary conditions can then be expressed as 
\begin{align}
    &v^{(b)}_x|_{y=+1} = \frac{v_u}{v_u - v_l} = \frac{\mathrm{Re}_u}{\mathrm{Re}_u - \mathrm{Re}_l}, &\text{and}&
    &v^{(b)}_x|_{y=-1} = \frac{v_l}{v_u - v_l} = \frac{\mathrm{Re}_l}{\mathrm{Re}_u - \mathrm{Re}_l}.
\end{align}
Using these in conjunction with (\ref{Eqn3.7}) we could obtain the exact expression for the velocity profile as an affine function,
\begin{align}
    v_x^{(b)}(y) = \frac{1}{2}\left(y + \frac{\mathrm{Re}_u +\mathrm{Re}_l}{\mathrm{Re}_u - \mathrm{Re}_l} \right). 
\end{align}
In order to find the solution for stress, we need to find the roots of the following expression 
\begin{align}\label{4.12b}
    F(c_1)=4\left(a + \left(b + 2\Gamma \mathrm{Re}^2c_1^2\right)^n\right)c_1 = 1.
\end{align}
Depending on the material parameters, one could find anywhere between one to three roots, which we shall do numerically. We shall assess the stability of these solutions in a following section. 
\subsubsection{Case 2: Mixed Boundary Conditions}\label{BCS}
Let us now consider the case where the value of stress is prescribed on one of the boundaries while we prescribe velocity on the other. Without loss of generality, we assume that the stress $s_u$ is applied on the upper plane and the lower plane slides with a constant velocity $v_b$. Using which we define a Reynolds number associated with the upper and lower planes as 
\begin{align}\label{4.13}
    &\mathrm{Re}_u = \rho\alpha^2 h^2 s_u, &\text{and}& &\mathrm{Re}_l = \rho\alpha h v_l.
\end{align}
The characteristic velocity scale $v^*$ in this case is then defined as $(\alpha d s_u - v_l)$ and therefore the characteristic stress becomes 
\begin{align}
    s^* = \frac{v^*}{\alpha h} = \frac{\alpha h s_u - v_l}{\alpha h}
\end{align}
we retain the definition of Reynolds number as before
\begin{align}
    &\mathrm{Re} = \rho\alpha h (\alpha h s_u -v_l) = \mathrm{Re}_u - \mathrm{Re}_l.
\end{align}
The non-dimensional boundary conditions are
\begin{align}
    &s^{(b)}_{xy}|_{y=+1} = \frac{s_u}{s^*} = \frac{\alpha h s_u}{\alpha h s_u - v_l} = \frac{\mathrm{Re}_u}{\mathrm{Re}_u - \mathrm{Re}_l}, &\text{and}&
    &v^{(b)}_x|_{y=-1} = \frac{v_l}{v_u - v_l} = \frac{\mathrm{Re}_l}{\mathrm{Re}_u - \mathrm{Re}_l}.
\end{align}
We could now use these along with (\ref{Eqn3.6}) and (\ref{Eqn3.7}) to determine the particular solution as
\begin{align}
    s^{(b)}_{xy}(y) &= \frac{\mathrm{Re}_u}{\mathrm{Re}_u -\mathrm{Re}_l}\\
    v^{(b)}_x(y) &= 2\left(a + \left(b + 2\Gamma \mathrm{Re}_u^2\right)^n\right)\frac{\mathrm{Re}_u}{\mathrm{Re}_u -\mathrm{Re}_l}(y+1) + \frac{\mathrm{Re}_l}{\mathrm{Re}_u -\mathrm{Re}_l}. 
\end{align}
\section{Stability}
\noindent Now let us assume that the perturbed flow is given by the following ansatz - 
\begin{align}
    &\mathbf{v} = \mathbf{v}^{(b)} + \mathbf{v}^{(p)} = \begin{bmatrix}
        v^{(b)}_x(y)\\
        0
    \end{bmatrix} + \begin{bmatrix}
        v^{(p)}_x(y)\\
        v^{(p)}_y(y)
    \end{bmatrix}e^{i(kx - st)}\\
    &\mathbf{S} = \mathbf{S}^{(b)} + \mathbf{S}^{(p)}  = \begin{bmatrix}
        0 & s^{(b)}_{xy}(y)\\
        s^{(b)}_{xy}(y) & 0
    \end{bmatrix} + \begin{bmatrix}
        s^{(p)}_{xx}(y) & s^{(p)}_{xy}(y)\\
        s^{(p)}_{xy}(y) & s^{(p)}_{yy}(y)
    \end{bmatrix}e^{i(kx - st)}\\  
    &p = p^{(b)} + p^{(p)}(y)e^{i(kx - st)}
\end{align}

\noindent The governing equations for the perturbed flow can be expressed as - 
\begin{subequations}
\begin{align}
&\text{div}(\mathbf{v}^{(p)}) = 0\\
&\mathrm{Re} \frac{D\mathbf{v}^{(p)}}{Dt} = -\text{grad}(p^{(p)}) + \text{div}(\mathbf{S}^{(p)})\\
&\mathbf{D}^{(b)} + \mathbf{D}^{(p)} = \left(a + \left(b + \beta \|\mathbf{S}^{(b)} + \mathbf{S}^{(p)}\|^2\right)^n\right)(\mathbf{S}^{(b)} + \mathbf{S}^{(p)})
\end{align}
\end{subequations}

\noindent Assuming that the perturbations are of infinitesimal amplitude, above equations can be linearized to obtain - 
\begin{align}
    &\left(a + \left(b + 2\Gamma \mathrm{Re}^2 {s^{(b)}_{xy}}^2\right)^n\right)s^{(p)}_{xx} = ikv^{(p)}_x\\
    &\left(a + \left(b + 2\Gamma \mathrm{Re}^2 {s^{(b)}_{xy}}^2\right)^n + 4n\Gamma \mathrm{Re}^2\left(b + 2\Gamma \mathrm{Re}^2 {s^{(b)}_{xy}}^2\right)^{n-1}{s^{(b)}_{xy}}^2 \right)s^{(p)}_{xy} = \frac{\left(v^{(p)}_{x,y}+ikv^{(p)}_y\right)}{2}\\
    &\left(a + \left(b + 2\Gamma \mathrm{Re}^2 {s^{(b)}_{xy}}^2\right)^n\right)s^{(p)}_{yy} = v^{(p)}_{y,y}\\
    &\mathrm{Re}\left(-isv^{(p)}_x + ikv^{(p)}_x v^{(b)}_x + v^{(p)}_y v^{(b)}_{x,y}\right) = -ikp^{(p)} + iks^{(p)}_{xx} + s^{(p)}_{xy,y}\label{Eqn3.29}\\
    &\mathrm{Re}\left(-isv^{(p)}_y + ikv^{(b)}_xv^{(p)}_y\right) = -p^{(p)}_{,y} + iks^{(p)}_{xy} + s^{(p)}_{yy,y} \label{Eqn3.30}\\
    &ikv^{(p)}_x + v^{(p)}_{y,y} = 0 \label{Eqn3.31}
\end{align}

\noindent In order to study the temporal stability of the above system of equations, we assume that $k\in \mathbb{R}$ and $s\in\mathbb{C}$. Since we already have our basic solution for $s^{(b)}_{xy}$ and $v^{(b)}_{x}$ from (\ref{Eqn3.6}) and (\ref{Eqn3.7}), we shall define two constants
\begin{align}
    &c_3 = \left(a + \left(b + 2\Gamma \mathrm{Re}^2c_1^2 \right)^n\right)\\
    &c_4 = \left(a + \left(b + 2\Gamma \mathrm{Re}^2c_1^2\right)^n + 4n\Gamma \mathrm{Re}^2\left(b + 2\Gamma \mathrm{Re}^2 c_1^2\right)^{n-1}c_1^2 \right)
\end{align}
We can then express 
\begin{align}
    &s^{(p)}_{xx} = \frac{ikv^{(p)}_x}{c_3}\\
    &s^{(p)}_{xy} = \frac{\left(v^{(p)}_{x,y}+ikv^{(p)}_y\right)}{2c_4}\\
    &s^{(p)}_{yy} = \frac{v^{(p)}_{y,y}}{c_3}
\end{align}
Upon substituting in (\ref{Eqn3.29}) and (\ref{Eqn3.30}),
\begin{align}
    &\mathrm{Re}\left(-isv^{(p)}_x + ikv^{(p)}_x v^{(b)}_x + v^{(p)}_y v^{(b)}_{x,y}\right) = -ikp^{(p)} - \frac{k^2v^{(p)}_x}{c_3} + \frac{\left(v^{(p)}_{x,yy}+ikv^{(p)}_{y,y}\right)}{2c_4} \label{Eqn3.36}\\
    &\mathrm{Re}\left(-isv^{(p)}_y + ikv^{(b)}_xv^{(p)}_y\right) = -p^{(p)}_{,y} + \frac{\left(ikv^{(p)}_{x,y}-k^2v^{(p)}_y\right)}{2c_4} + \frac{v^{(p)}_{y,yy}}{c_3} \label{Eqn3.37}
\end{align}
Multiplying (\ref{Eqn3.37}) with $-ik$ and adding to the $y$--derivative of (\ref{Eqn3.36}), we obtain 
\begin{align}
&\mathrm{Re}\left((ik v_x^{(b)}-is)v_{x,y}^{(p)} + (k^2 v_x^{(b)}-ks)v_y^{(p)}\right) = \frac{1}{2c_4}v_{x,{y}^{(3)}}^{(p)} 
+ \left(\frac{k^2}{c_4}-\frac{2k^2}{c_3}\right) v_{x,y}^{(p)}
+ \frac{ik^3}{2c_4}v_y^{(p)}. \label{Eqn3.38}
\end{align}
Using the continuity relation (\ref{Eqn3.31}), $ik\,v_x^{(p)} + v_{y,y}^{(p)}=0$, we can eliminate $v_x^{(p)}$ and write (\ref{Eqn3.38}) entirely in terms of $v_y^{(p)}$ as
\begin{align}\label{Eqn3.39}
\mathrm{Re}\left(\frac{s-kv_x^{(b)}}{k}\,v_{y,yy}^{(p)} + (k^2 v_x^{(b)}-ks)\,v_y^{(p)}\right) = \frac{i}{2c_4k}v_{y,{y}^{(4)}}^{(p)}
- i\left(\frac{2k}{c_3}-\frac{k}{c_4}\right) v_{y,yy}^{(p)} + \frac{ik^3}{2c_4}\,v_y^{(p)}. 
\end{align}
Introducing $s = k m$, where $m \in \mathbb{C}$, equation \eqref{Eqn3.39} becomes
\begin{align}
\mathrm{Re}(v_x^{(b)}-m)\left(k^2 v_y^{(p)}-v_{y,yy}^{(p)}\right)= \frac{i}{2c_4k}\,v_{y,{y}^{(4)}}^{(p)} - i\left(\frac{2k}{c_3}-\frac{k}{c_4}\right) v_{y,yy}^{(p)}
+ \frac{ik^3}{2c_4}\,v_y^{(p)} .
\label{3.40}
\end{align}
Multiplying through by $ik$ yields the Orr--Sommerfeld--type form
\begin{align}
ik\mathrm{Re}(v_x^{(b)}-m)(k^2 v_y^{(p)} - v_{y,yy}^{(p)})= -\frac{1}{2c_4}\,v_{y,{y}^{(4)}}^{(p)}+ \left(\frac{2k^2}{c_3}-\frac{k^2}{c_4}\right) v_{y,yy}^{(p)}- \frac{k^4}{2c_4}\,v_y^{(p)}, \label{3.41}
\end{align}
which could then be expressed as an eigenvalue problem
\begin{align}
\big(A_0 + m A_1\big)\,v_y^{(p)} = 0,
\label{3.42}
\end{align}
using the differential operator $\mathrm{D}$
\begin{align}
A_0 &= A_{00} \mathrm{I} + A_{01} \mathrm{D} + A_{02} \mathrm{D}^2 + A_{03} \mathrm{D}^3 + A_{04} \mathrm{D}^4, \notag \\
A_1 &= A_{10} \mathrm{I} + A_{11} \mathrm{D} + A_{12} \mathrm{D}^2,
\label{3.43}
\end{align}
where the coefficients are
\begin{align*}
&A_{00}=-ik^3\mathrm{Re}\,v_x^{(b)}+\frac{k^4}{2c_4}, \quad A_{01}=0, \quad A_{02}=ik\,\mathrm{Re}\,v_x^{(b)} - \frac{2k^2}{c_3}+\frac{k^2}{c_4}, \quad A_{03}=0, \quad A_{04}=\frac{1}{2c_4},\\
&A_{10}=ik^3 \mathrm{Re}, \quad A_{11}=0, \quad A_{12}=-ik\,\mathrm{Re}.
\end{align*}
Once the eigenvalue problem \eqref{3.42} has been solved, we can write the perturbation of the transverse component of the velocity as
\begin{align}
\tilde{v}_y(x,y,t)=v_y^{(p)}(y)e^{ik(x-mt)}.
\end{align}
To determine the longitudinal component
\begin{align}
\tilde{v}_x(x,y,t)=v_x^{(p)}(y)e^{ik(x-mt)}.
\end{align}
we observe that, from the mass balance,
\begin{align}
\tilde{v}_x(x,y,t)=\dfrac{i\tilde{v}_{y,y}(x,y,t)}{k},
\end{align}
so that 
\begin{align}
v_x^{(p)}(y)=\dfrac{iv_{y,y}^{(p)}(y)}{k}.
\end{align}
We then introduce the complex stream function 
\begin{align}
\tilde{\psi}(x,y,t)=\psi^{(p)}(y)e^{ik(x-mt)},
\end{align}
defined so that
\begin{align}
\tilde{v}_x(x,y,t)=\tilde{\psi}_{,y}(x,y,t), \ \ \ \ \ \ 
\tilde{v}_y(x,y,t)=-\tilde{\psi}_{,x}(x,y,t). 
\end{align}
It is easily found by integration that
\begin{align}\label{stream}
\tilde{\psi}(x,y,t)=\dfrac{i}{k}v_{y,y}^{(p)}(y)e^{ik(x-mt)}+const.
\end{align}
The curves $\mathfrak{R}(\tilde{\psi}(x,y,t))=const$ represent the streamlines of the perturbation. The spatial period of the streamlines is $2\pi/k$. 

\section{Numerical Method}

In the eigenvalue problem \eqref{3.42}  \(A_0\) and \(A_1\) are linear differential operators, $v_y^{(p)}(y)$  is the eigenfunction, and \(m \in \mathbb{C}\) denotes the complex eigenvalue. The problem is solved numerically using a pseudospectral collocation method based on Gauss--Lobatto points.

The computational domain is discretized by \(N+1\) collocation points \(\{y_j\}_{j=0}^N\), corresponding to the Gauss--Lobatto nodes associated with Legendre (or Chebyshev) polynomials. The continuous function $v_y^{(p)}(y)$ is approximated by an interpolating polynomial,
\begin{equation}
v_y^{(p)}(y) \approx \sum_{j=0}^{N} v_y^{(p)}(y_j) \, \ell_j(y),
\end{equation}
where \(\ell_j(y)\) are the Lagrange cardinal polynomials satisfying \(\ell_j(y_i) = \delta_{ij}\). The derivatives of $v_y^{(p)}(y)$ at the collocation points are computed by means of the \emph{differentiation matrix} \(D\), whose entries are given by \(D_{ij} = \ell_j'(y_i)\). Higher-order derivatives are obtained recursively as \(D^{(k)} = D^k\). 
Applying the pseudospectral discretization to the differential operators  \(A_0\) and \(A_1\) yields the discrete matrices \(A_0^{(N)}\) and \(A_1^{(N)}\), which act on the vector of nodal values
\[
\mathbf{v} = [v_0, v_1, \dots, v_N]^T, \ \ \ \  v_i = v_y^{(p)}(y_i).
\]
Boundary conditions are imposed by appropriately modifying the rows of the matrices or by eliminating the boundary degrees of freedom. The resulting discrete eigenvalue problem reads
\begin{equation}
\left( A_0^{(N)} + m A_1^{(N)} \right) \mathbf{v} = \mathbf{0},
\end{equation}
which is an algebraic eigenvalue problem. 
This system is then solved using standard numerical eigensolvers (e.g., the QZ algorithm) to obtain the discrete set of eigenvalues \(m_k\) and corresponding eigenvectors \(\mathbf{v}_k\).

Convergence of the pseudospectral discretization is spectral for sufficiently smooth eigenfunctions, meaning that the error decreases exponentially with increasing \(N\). The use of Gauss--Lobatto nodes ensures that the collocation points include the domain boundaries, facilitating the accurate imposition of boundary conditions.

\section{Results}

In this section, we present some stability results obtained from the solution of the eigenvalue problem \eqref{3.42}. The spectrum of the discretized operator provides direct information on the linear stability of the system. In particular, the complex part of the computed eigenvalues determines whether small perturbations decay or grow in time. By analyzing the distribution of eigenvalues in the complex plane, we identify the stability characteristics of the base state and determine the conditions under which instability arises.

\subsection{Case 1: Velocity Boundary Conditions }

\begin{figure}
    \centering
    \includegraphics[width=0.6\linewidth]{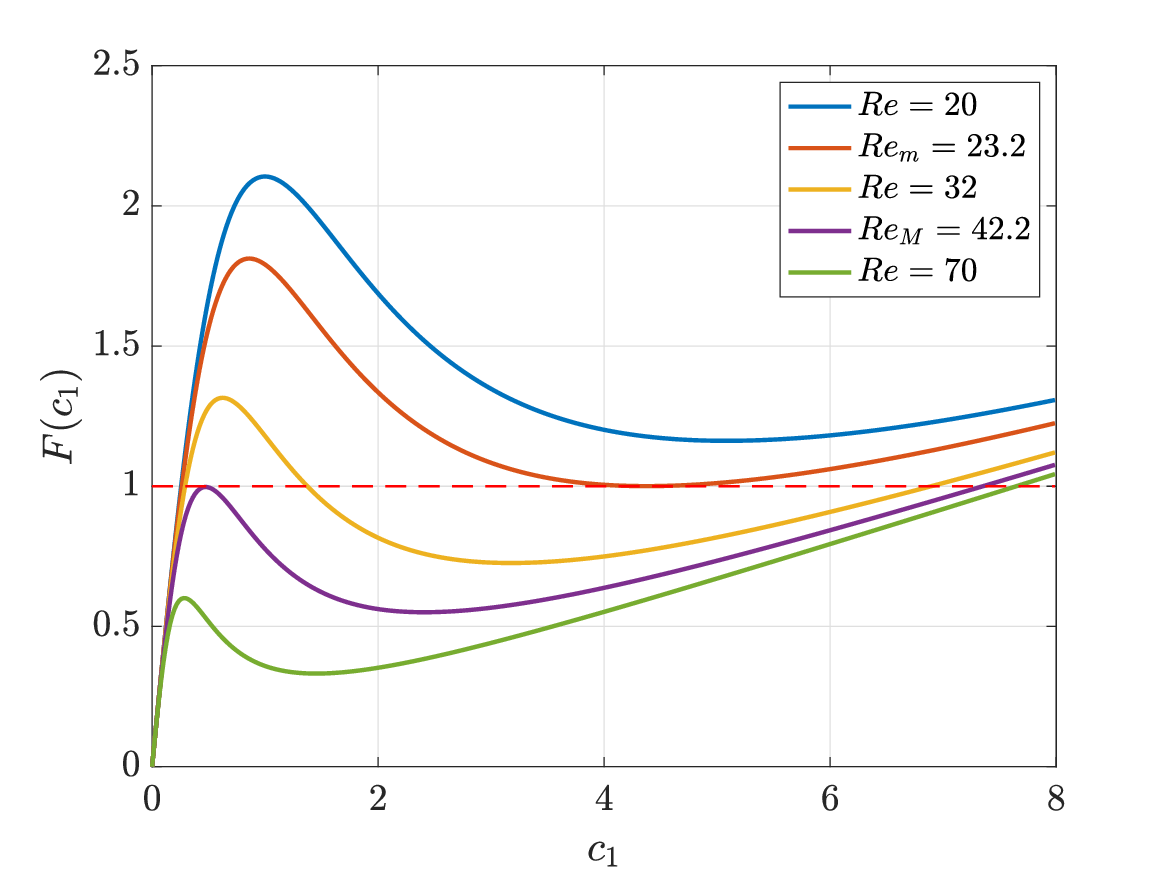}
    \caption{Plot of the function $F(c_1)$ defined in \eqref{4.12b} with $a=0.032$, $b=1$, $\Gamma=10^{-3}$, $n=-1.2$ and for various $Re$. The range in which 3 solutions occur is $(Re_m,Re_M)$, with $Re_m=23.2$ and $Re_M=42.2$.}
    \label{fig:01}
\end{figure}

Let us begin by considering the case in which the velocity is prescribed on the upper wall, see section \ref{BCV}. This is the setting in which, depending on the imposed velocities on the upper and lower walls, three different velocity profiles may coexist. We observe that, depending on the material parameters, there exists a limited range of $Re=Re_u-Re_l$ in which three basic solutions can occur. In fact, looking at Fig. \ref{fig:01}, which shows the function $F(c_1)$ defined in \eqref{4.12b}, it can be seen that, if $Re\in [Re_m,Re_M]$, there can be three solutions to equation \eqref{4.12b}, while outside this range there is only one. The material parameters used in the plots in Fig. \ref{fig:01} are $a=0.032$, $b=1$, $\Gamma=10^{-3}$, $n=-1.2$ (non-monotonic constitutive equation) and $Re_m=23$, $Re_M=42.2$.

\begin{figure}
    \centering
    \includegraphics[width=0.6\linewidth]{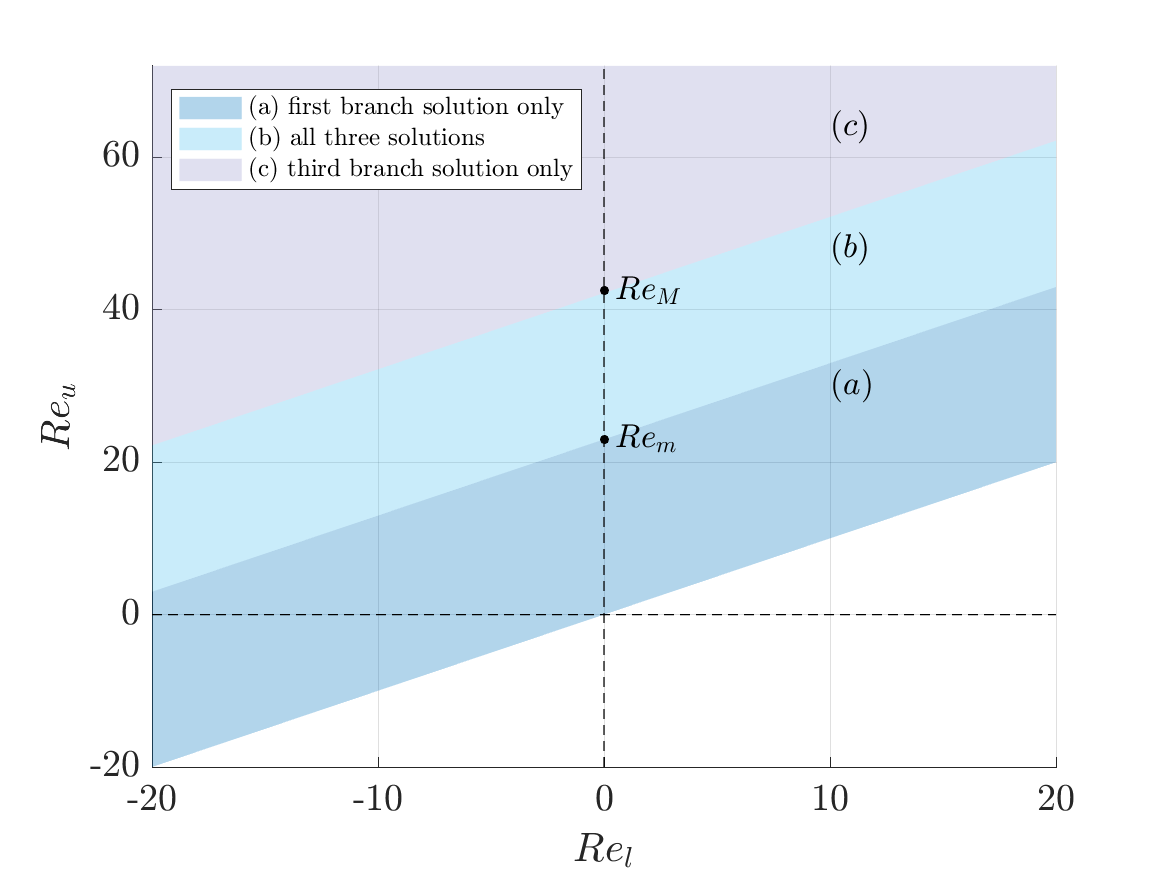}
    \caption{Regions of existence of the three solutions.}
    \label{fig:02}
\end{figure}

In Fig. \ref{fig:02} we show, in the $(Re_l, Re_u)$ plane, the regions of interest for studying the stability of the basic solutions. Without loss of generality, we can restrict ourselves to the portion of the plane where $Re_u \geqslant Re_l$, i.e. where $Re \geqslant 0$ (the case $Re \leqslant 0$ can be obtained by symmetry). In this part of the plane, for a non-monotonic constitutive equation, we observe three distinct zones in which we have: $(a)$ existence of only one solution (stress corresponding to the first ascending branch of the constitutive law); $(b)$ existence of three solutions (stresses corresponding to the two ascending branches and to the descending branch of the constitutive law); $(c)$ existence of only one solution (stress corresponding to the second ascending branch of the constitutive law). From Figs. \ref{fig:01} and \ref{fig:02}, it is evident that the solution corresponding to the first ascending branch  can exist only for $(Re_l, Re_u)$ belonging to the region $(a) \cup (b)$. On the other hand, the solution corresponding to the third ascending branch can exist only for $(Re_l, Re_u) \in (b) \cup (c)$. Finally, the solution corresponding to the descending branch can exist only for $(Re_l, Re_u) \in (b)$.

To study the stability of the basic solutions in regions $(a)$, $(b)$, and $(c)$, we numerically solve the eigenvalue problem \eqref{3.42} for each pair of values $(Re_l,Re_u)$ belonging to the aforementioned regions. In particular, we use the basic solution corresponding to the first increasing branch in region $(a)\cup(b)$, the basic solution corresponding to the decreasing branch in $(b)$, and the basic solution corresponding to the second increasing branch in region $(b)\cup(c)$. For simplicity, we denote these solutions as follows: 
\begin{align}
    &v^{(b,1)}_{x}(y)\,\, \textrm{first ascending branch}\,\,\, (a) \cup (b), \\
    &v^{(b,2)}_{x}(y)\,\, \textrm{descending branch}\,\,\, (b), \\
    &v^{(b,3)}_{x}(y)\,\, \textrm{second ascending branch}\,\,\, (a) \cup (b).
\end{align}
For each basic solution we consider the eigenvalue with the largest imaginary part and define such imaginary part as $m_i(Re_l,Re_u;k;v^{(b,i)}_{x})$, $ (i=1,2,3)$. This is a real-valued function of $Re_l$, $Re_u$, $k$, $v^{(b,i)}_{x}$ plus all the material parameters appearing in the problem. We plot the function $m_i$ in the plane $(Re_l,Re_u)$, more precisely in the region in which the basic solution 
$v^{(b,i)}_{x}$ is defined. We assume that the wavenumber $k$ is fixed and equal to one (we shall see that stability results are not altered for different values of $k$). 

\begin{figure}[ht!]
    \centering
    \captionsetup[subfigure]{labelformat=empty}

    \begin{subfigure}{0.49\linewidth}
        \centering
        \includegraphics[width=\linewidth,trim={0.55cm 0.2cm 0.55cm 0.2cm},clip]{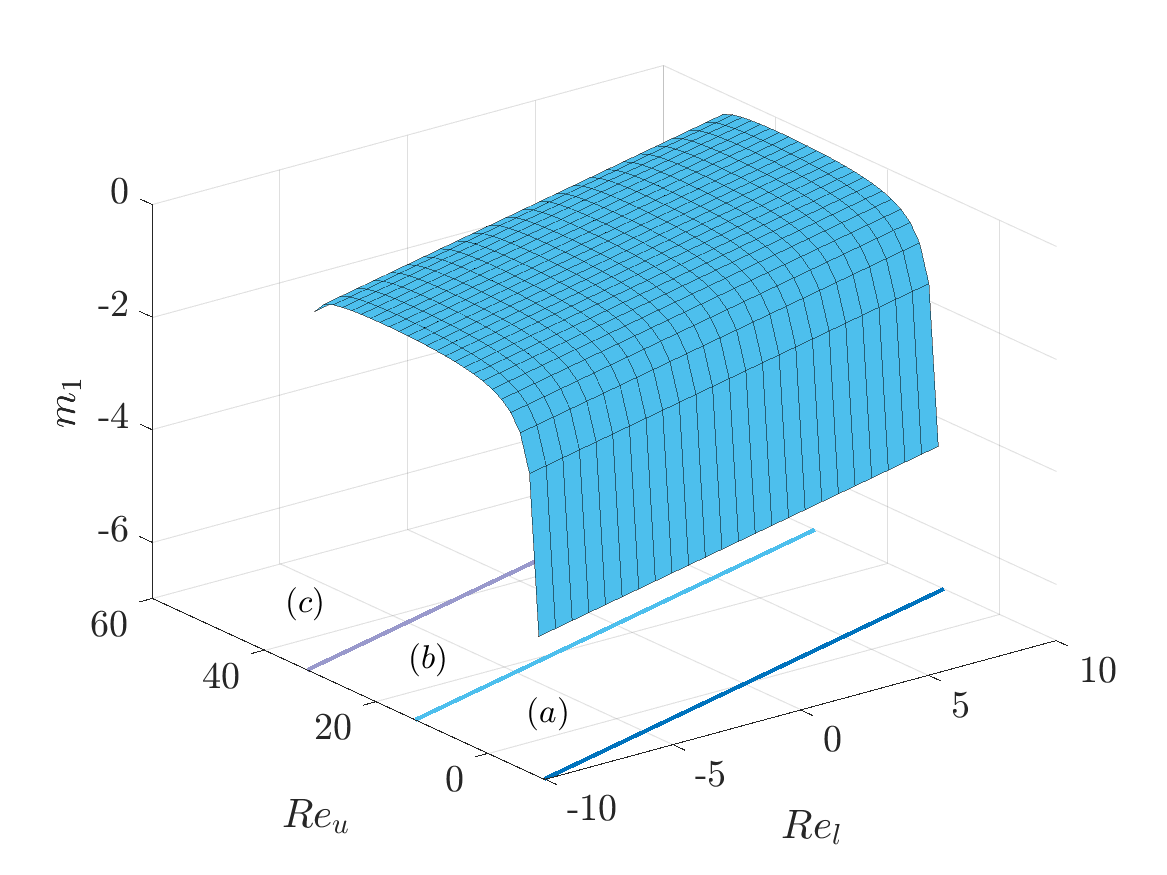}
        \caption{(i) $m_1(Re_l,Re_u)$ for regions $(a)\cup(b)$.}
        \label{fig:03}
    \end{subfigure}
    \hfill
    \begin{subfigure}{0.49\linewidth}
        \centering
        \includegraphics[width=\linewidth,trim={0.55cm 0.2cm 0.55cm 0.2cm},clip]{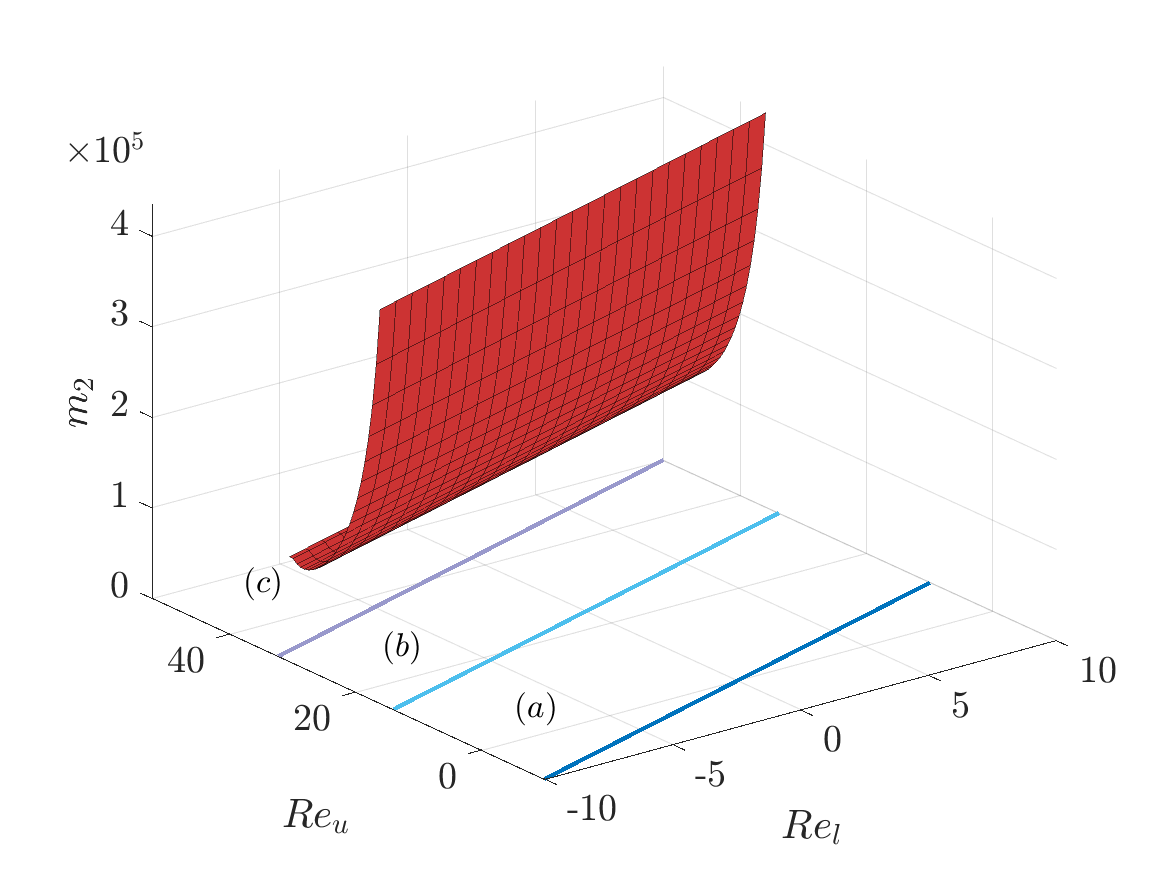}
        \caption{(ii) $m_2(Re_l,Re_u)$ for region $(b)$.}
        \label{fig:04}
    \end{subfigure}

    \vspace{0.9em}

    \begin{subfigure}{0.5\linewidth}
        \centering
        \includegraphics[width=\linewidth,trim={0.55cm 0.1cm 0.55cm 0.2cm},clip]{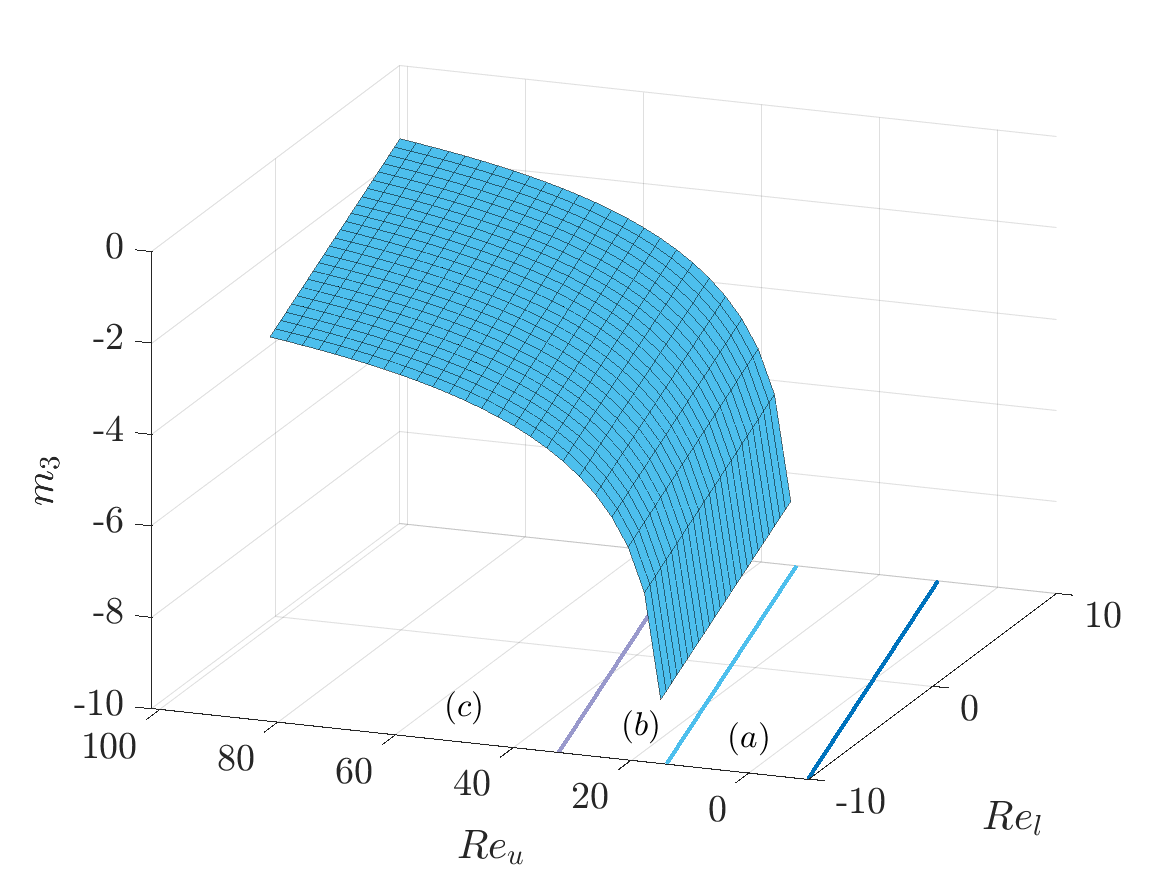}
        \caption{(iii) $m_3(Re_l,Re_u)$ for regions $(b)\cup(c)$.}
        \label{fig:05}
    \end{subfigure}

    \caption{
        $m_i(Re_l,Re_u)$ corresponding to the three base
        states $v_x^{(b,i)}(y)$, $i=1,2,3$.  
        States (i) and (iii), lying on the ascending branches of the constitutive
        curve, are unconditionally stable, whereas state (ii), associated with the
        descending branch, is unconditionally unstable.}
    \label{fig:triangle_stability}
\end{figure}

In Figs. \ref{fig:03}, \ref{fig:04}, \ref{fig:05}, we show the plots of the function $m_i(Re_l,Re_u;k;v^{(b,i)}_{x})$ corresponding to the three solutions $v^{(b,i)}_{x}(y)$, $i=1,2,3$ in the regions where they are defined. We observe that solutions $v^{(b,1)}_{x}(y)$,$v^{(b,3)}_{x}(y)$ are unconditionally stable ($m_1, m_3<0,\,$ Figs. \ref{fig:03}, \ref{fig:05}) in regions $(a)\cup(b)$ and $(b)\cup(c)$, respectively, while solution $v^{(b,2)}_{x}(y)$ is unconditionally unstable ($m_2>0$, Fig. \ref{fig:04}) in $(b)$. To better highlight the region where the functions $m_i$ are defined, in each plot we have shown the projection of the three definition domains $(a)$, $(b)$ and $(c)$.

From the plots in Figs. \ref{fig:03}, \ref{fig:04}, \ref{fig:05}, it is  evident that the functions $m_i(Re_l,Re_u;k;v^{(b,i)}_{x})$ are constant along  the lines $Re_u-Re_l=const$, that is the function $m_i$ depends only on $Re=Re_u-Re_l$, i.e. $m_i(Re;k;v^{(b,i)}_{x})$. Stability/instability is thus determined (when Dirichlet boundary connditions for the velocity are applied) by the relative velocity of the upper and lower plate, not by their absolute values. 

To investigate the dependence of $m_i$ on the wavenumber $k$ we can also study the function $m_i(Re;k;v^{(b,i)}_{x})$ relative to solution $v^{(b,i)}_{x}(y)$, for the admissible values of $Re$. We find that for any $k$ the solutions $v^{(b,1)}_{x}(y)$,$v^{(b,3)}_{x}(y)$ are unconditionally stable $(m_1, m_3<0)$, whereas $v^{(b,2)}_{x}(y)$ is unconditionally unstable $(m_2>0)$, so that stability is not influenced by the particular wavenumber. 

To better study the behavior of the functions $m_1$  and $m_3$ with respect to the Reynolds number $Re$, we can plot the curves $m_1$, $m_3$ for fixed values of $k$.

\begin{figure}[ht!]
    \centering
    \captionsetup[subfigure]{labelformat=empty}

    \begin{subfigure}{0.49\linewidth}
        \centering
        \includegraphics[width=\linewidth,trim={0.8cm 0.2cm 0.8cm 0.2cm},clip]{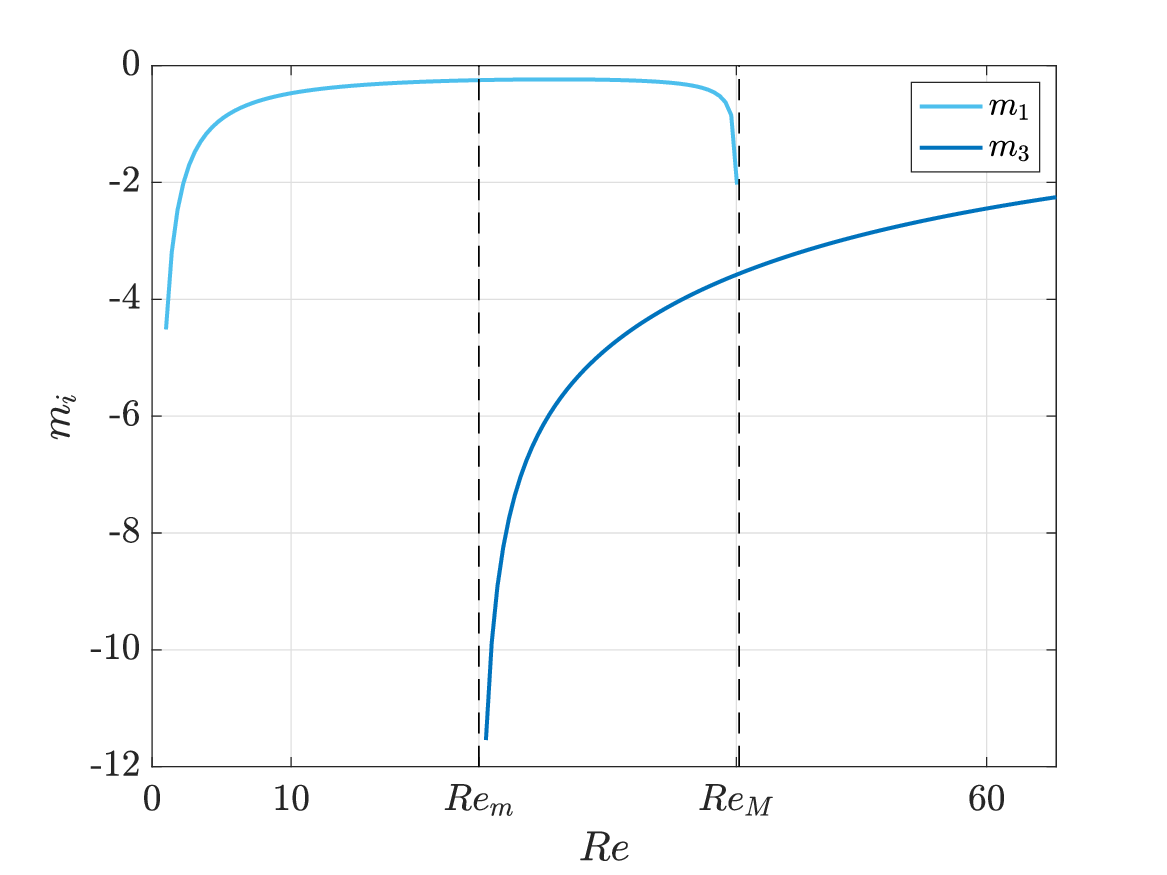}
        \caption{(i) $m_1$ and $m_3$ versus $Re$ for fixed $k=1$.}
        \label{fig:04bis}
    \end{subfigure}
    \hfill
    \begin{subfigure}{0.49\linewidth}
        \centering
        \includegraphics[width=\linewidth,trim={0.8cm 0.2cm 0.8cm 0.2cm},clip]{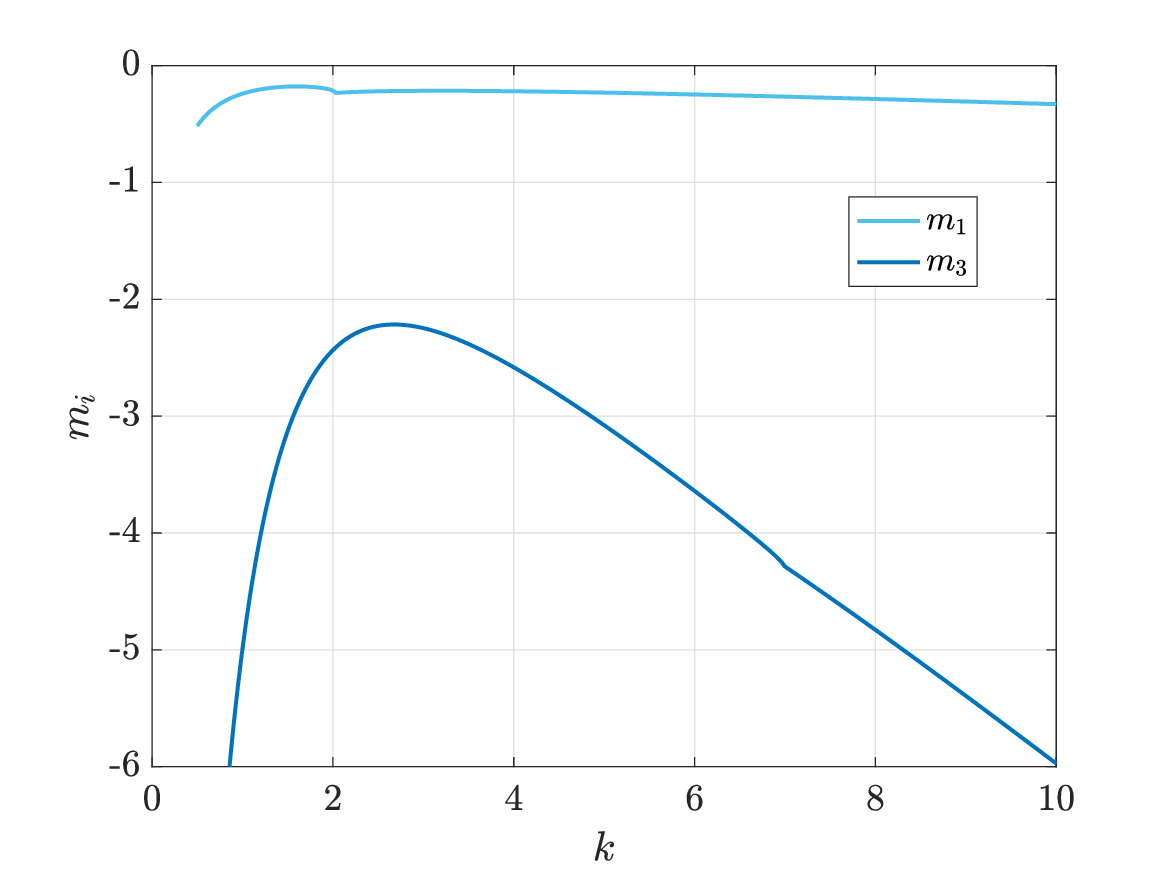}
        \caption{(ii) $m_1$ and $m_3$ versus $k$ for fixed $Re=32$.}
        \label{fig:04ter}
    \end{subfigure}

    \caption{Plot of the functions $m_1$, $m_3$ versus (i) $Re$ with $k=1$ and (ii) $k$ with $Re=32$.}
    \label{fig:merged_m1m3}
\end{figure}

The curves in Fig. \ref{fig:04bis} represent the functions $m_i,\, i = 1,3$ in terms of $Re$ for $k=1$. The region between the two vertical dashed lines indicates the coexistence zone of the stable solutions 1 and 3. It is observed that the function $m_1$ does not tend to zero as $Re\rightarrow Re_M^-$. The same behavior occurs for the function  $m_3$ as $Re\rightarrow Re_m^+$ (both limits are constant values, not infinite). This implies that the demarcation lines between regions (a), (b), and (c)
do not represent ``true'' marginal stability curves, since the function $m_i$ does not vanish on those lines. We also note that the in the region of coexistence $m_3$ is always smaller than $m_1$, meaning that the third solution is always more stable\footnote{By ``more stable'' we mean that the perturbation decays at a faster rate.} than the first one. This is in accordance with the constitutive behavior depicted in Fig. \ref{fig:01}, in which the slope of the first solution (yellow curve) is smaller than that of the third solution, indicating that the third solution is ``more viscous'' than the first one and hence more stable.  

\begin{figure}[ht!]
    \centering
    
    \begin{subfigure}{0.49\linewidth}
        \centering
        \includegraphics[width=\linewidth,trim={0.65cm 0.2cm 0.65cm 0.2cm},clip]{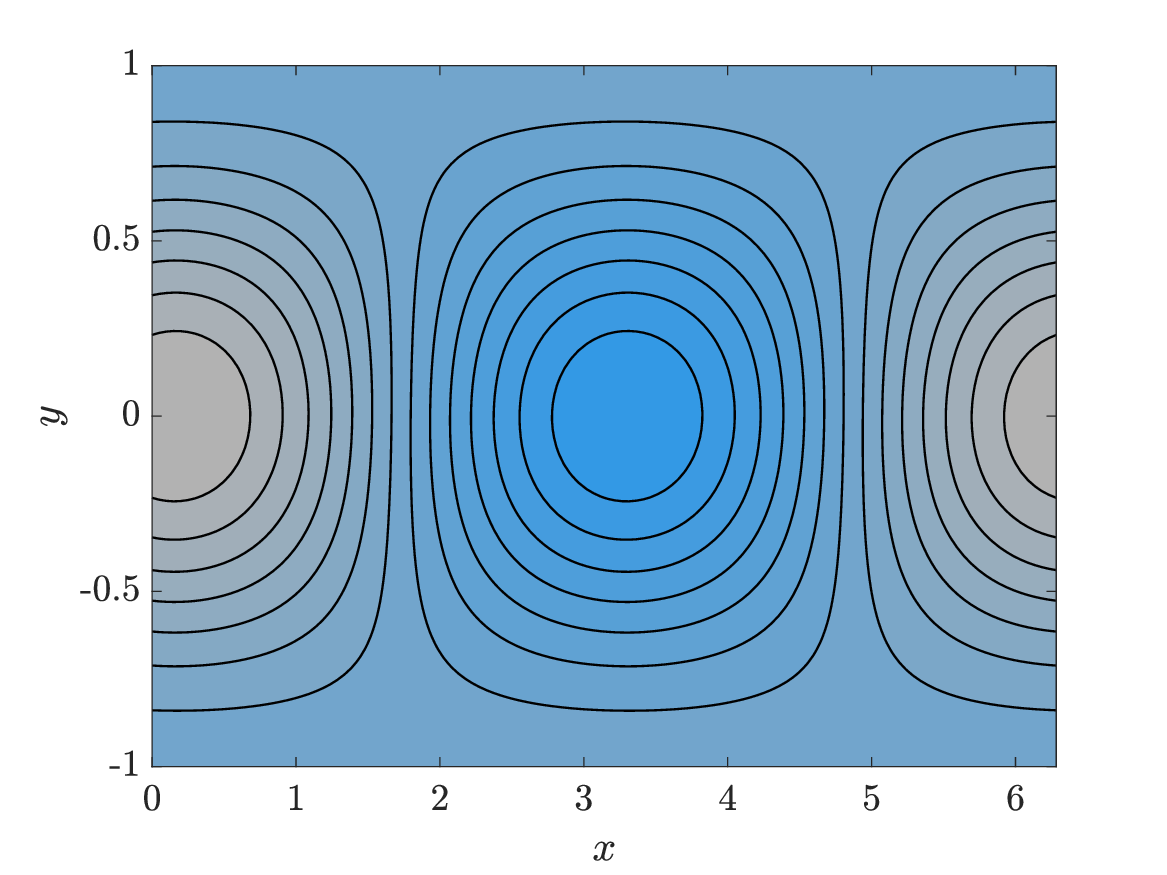}
        \caption{$k=1$, $Re=80$ (third solution).}
        \label{fig:04quater}
    \end{subfigure}
    \hfill
    \begin{subfigure}{0.49\linewidth}
        \centering
        \includegraphics[width=\linewidth,trim={0.65cm 0.2cm 0.65cm 0.2cm},clip]{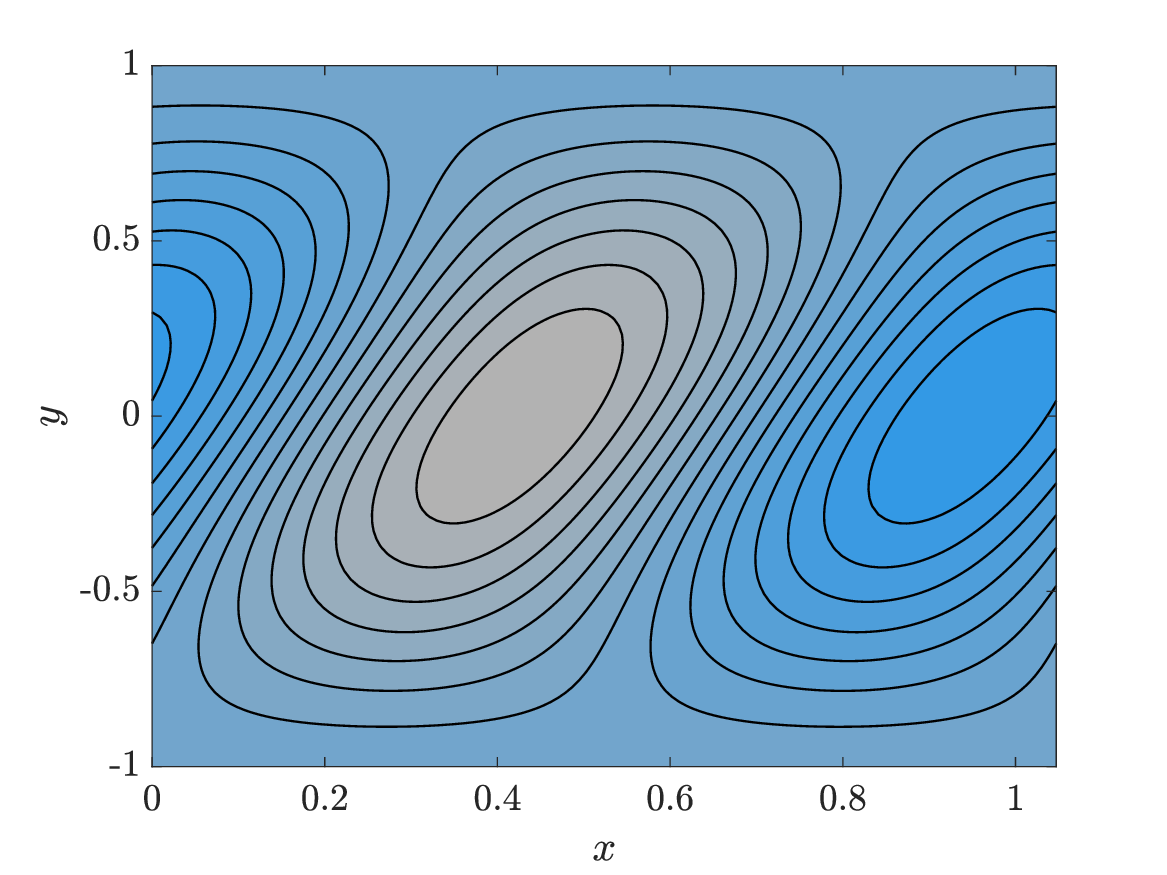}
        \caption{$k=6$, $Re=80$ (third solution).}
        \label{fig:04five}
    \end{subfigure}

    \caption{Streamlines $\mathfrak{R}(\tilde{\psi})=\mathrm{const}$ for two representative wavenumbers,
    computed for the third (stable) base state with parameters 
    $a=0.032$, $b=1$, $\Gamma=10^{-3}$, $n=-1.2$.}
    \label{fig:streamline_combined}
\end{figure}

In Fig. \ref{fig:04ter} we plot the functions $m_1,\, m_3$ vs $k$ for $Re=32$, i.e. for a Reynolds number for which we have the coexistence of both solutions. We again observe that the third solution is more stable than the first one. Finally, in Figs. \ref{fig:04quater}, \ref{fig:04five} we show the streamlines $\mathfrak{R}(\tilde{\psi})=const$ of the perturbation defined in \eqref{stream}  $k=1, 6$, $Re=80$ (third solution). We observe that the increase of $k$ results in a stronger shearing of the fluid.

\subsection{Case 2: Mixed Boundary Conditions }

Let us now consider the case in which a stress is prescribed on the upper wall, and velocity is imposed on the lower. The stress in the fluid layer is constant and equal to the one imposed on the upper wall. The solution is now uniquely determined, since a fixed stress corresponds to a single velocity profile.

Assigning the stress on the upper wall and the velocity on the lower wall is equivalent to fixing $Re_u$ and $Re_l$ (and therefore $Re=Re_u-Re_l$) defined as in \eqref{4.13}. In this case, there are no restrictions on $Re$, except for the assumption $Re \geqslant 0$ (the opposite case can be easily obtained by symmetry).

The prescribed stress on the upper wall is $Re_u/Re$ and corresponds to a single value of $v^{(b)}_{x,y}$, which may lie on either one of the two ascending branches or on the descending branch of the constitutive equation. In the first case, the basic solution is unconditionally stable, whereas in the second it is unconditionally unstable.

\begin{figure}[ht!]
    \centering
    \includegraphics[width=0.6\linewidth]{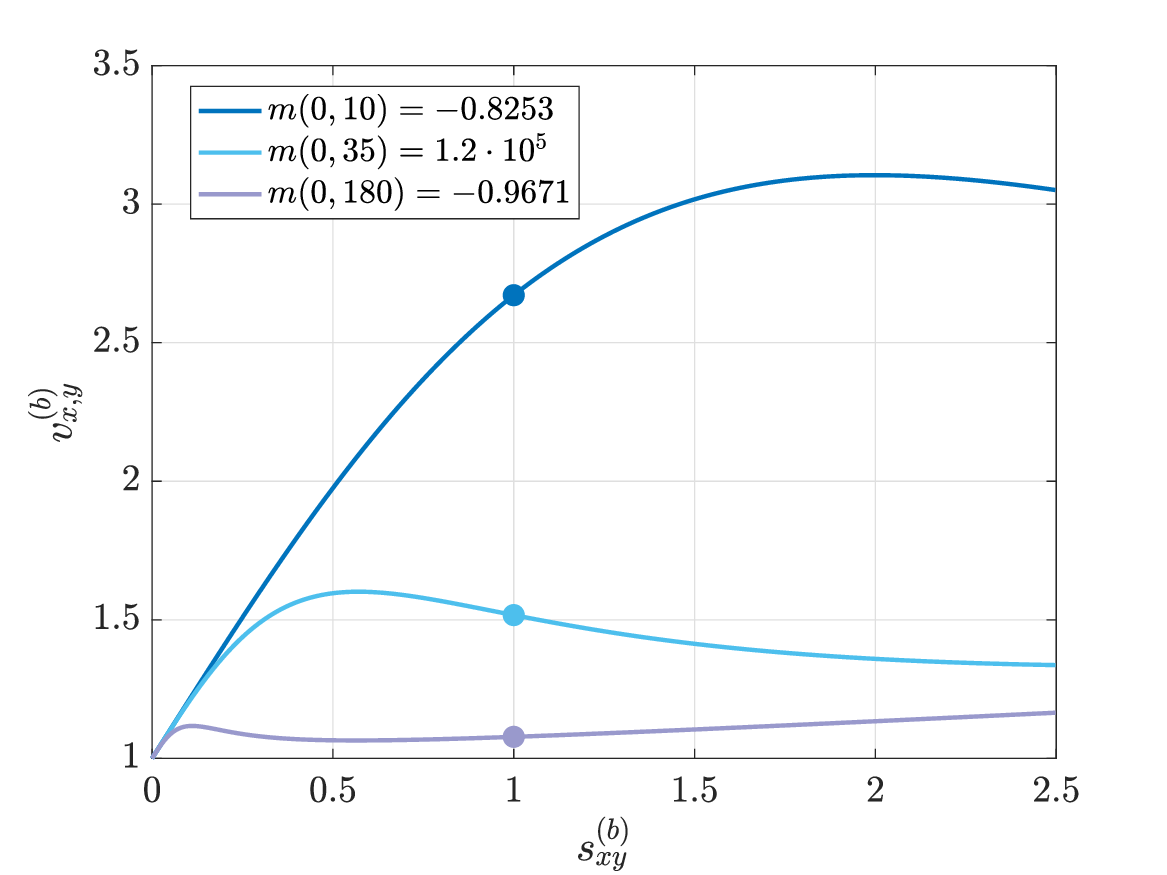}
    \caption{$v^{(b)}_{x,y}$ vs $s^{(b)}_{xy}$, eq. \eqref{4.4}. The material parameters are $n=-1.2$, $a=0.032$, $b=1$, $\Gamma=10^{-3}$. $Re_l=0$, $Re_u=10,\, 35,\, 180$. The coordinates of the dot represent the values of the shear stress and the corresponding velocity gradient.}
    \label{fig:09}
\end{figure}
In Fig. \ref{fig:09} we show the plots of the strain rate versus stress function \eqref{4.4}
for $Re_l=0$, $Re=Re_u=10, 35, 180$, $n=-1.2$, $a=0.032$, $b=1$, $\Gamma=10^{-3}$. The point on the curve corresponds to the stress $Re_u/Re$ applied on the top (in this case it is 1 because $Re_l=0$). It can be observed that, depending on the applied shear stress, this point may lie on either of the two ascending branches or on the descending one. In the former case, the function  $m(Re_l,Re_u;k;v^{(b)}_{x})$ (which is now uniquely defined, since the stress is prescribed) is strictly negative (we can easily check this solving the eigenvalue problem), indicating unconditional stability, whereas in the latter it is strictly positive, indicating unconditional instability.

\begin{figure}[ht!]
    \centering
    \includegraphics[width=0.6\linewidth]{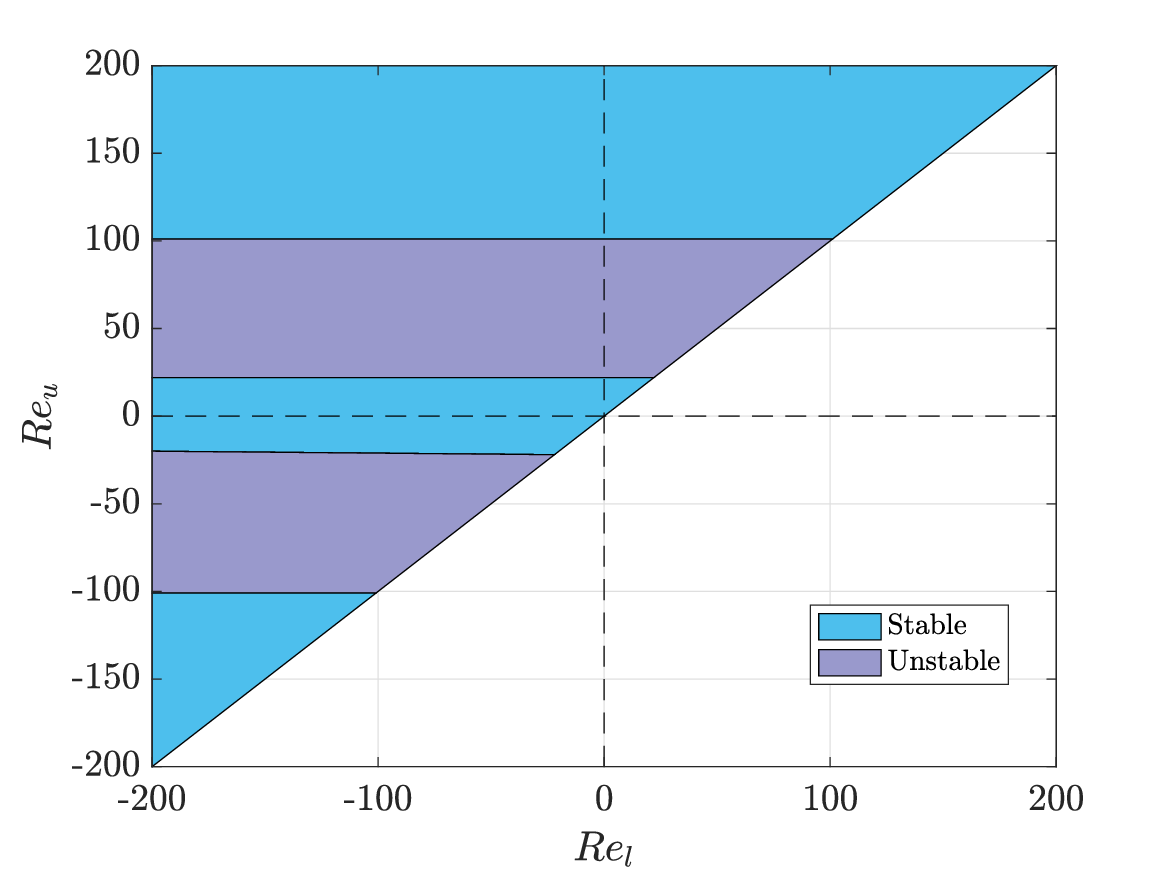}
    \caption{Stability diagram. Stress $Re_u/Re$ applied on the top surface.}
    \label{fig:10}
\end{figure}

In Fig. \ref{fig:10} we plot in the $Re_u\geqslant Re_l$ plane the marginal stability curves for the system with stress $Re_u/Re$ applied on the upper wall. The material parameters are the same of Fig. \ref{fig:09}. The blue area corresponds to the pairs $(Re_l,Re_u)$ in which the basic solution is stable, whereas the gray area corresponds to those in which the solution is unstable. We observe that the stability/instability regions are symmetric with respect to the line $Re_u = 0$, which allows us to limit our analysis to the portion $Re_u \geqslant 0$.

 From the figure it can be seen that, regardless of the velocity imposed on the lower wall, there exist two critical values $Re^1_u$, $Re^2_u$ ($Re_u^1=22$, $Re_u^2=101$) such that, for $Re_u>0$ outside the interval $[Re^1_u, Re^2_u]$, the basic solution is stable, whereas for $Re_u\in[Re^1_u, Re^2_u]$ the solution is unstable. Although the marginal stability curves are symmetric with respect to $Re_u=0$, the function $m(Re_l,Re_u;k;v^{(b)}_{x})$ is not, meaning that the decay rate  for a particular $(Re_l,Re_u)$ is different from that of $(Re_l,-Re_u)$.

\begin{figure}
    \centering
    \includegraphics[width=0.6\linewidth]{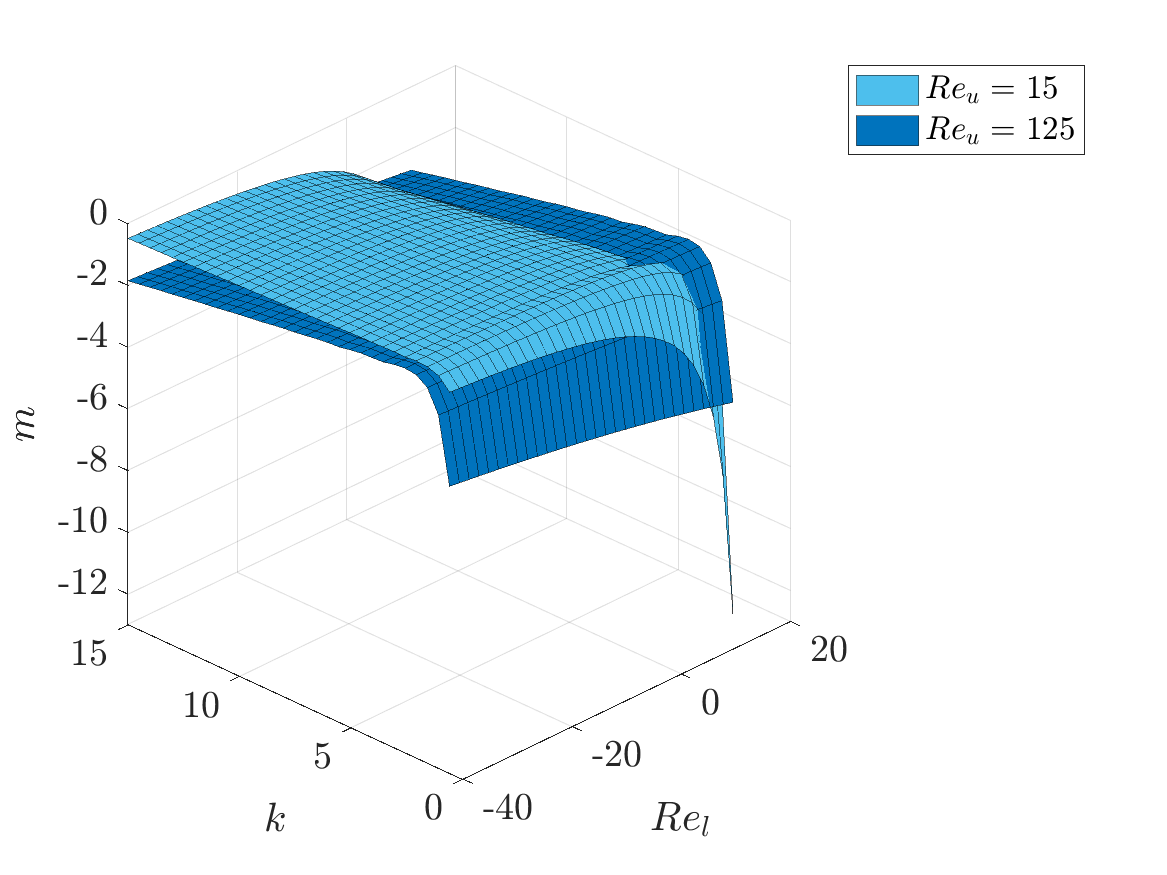}
    \caption{Function $m(Re_l,Re_u;k;v^{(b)}_{x})$ as a function of $Re_l$ and $k$ with $Re_u=15, 125$, $n=-1.2$, $a=0.032$, $b=1$, $\Gamma=10^{-3}$. }
    \label{fig:12}
\end{figure}

Fig. \ref{fig:12} shows the surface $m(Re_l,Re_u;k;v^{(b)}_{x})$ as a function of $Re_l$ and $k$ with $Re_u=15, 125$, $n=-1.2$, $a=0.032$, $b=1$, $\Gamma=10^{-3}$.  It is evident that 
$m$ is a non-increasing function of $Re_l$ and  exhibits a non-monotonic dependence on 
$k$. In particular, with regard to the non-monotonic behavior in $k$, it can be observed that the trend closely resembles that obtained in the case of a prescribed velocity at the upper wall (see Fig. \ref{fig:04ter}). Furthermore, we see that when the difference $Re=Re_u-Re_l$ becomes large, the solution corresponding to a larger value of $Re_u$ is more stable, but for small value of $Re$ the solution relative to the smaller $Re_u$ is more stable. 

\begin{figure}[ht!]
    \centering

    \begin{subfigure}{0.49\linewidth}
        \centering
        \includegraphics[width=\linewidth,trim={0.65cm 0.2cm 0.65cm 0.2cm},clip]{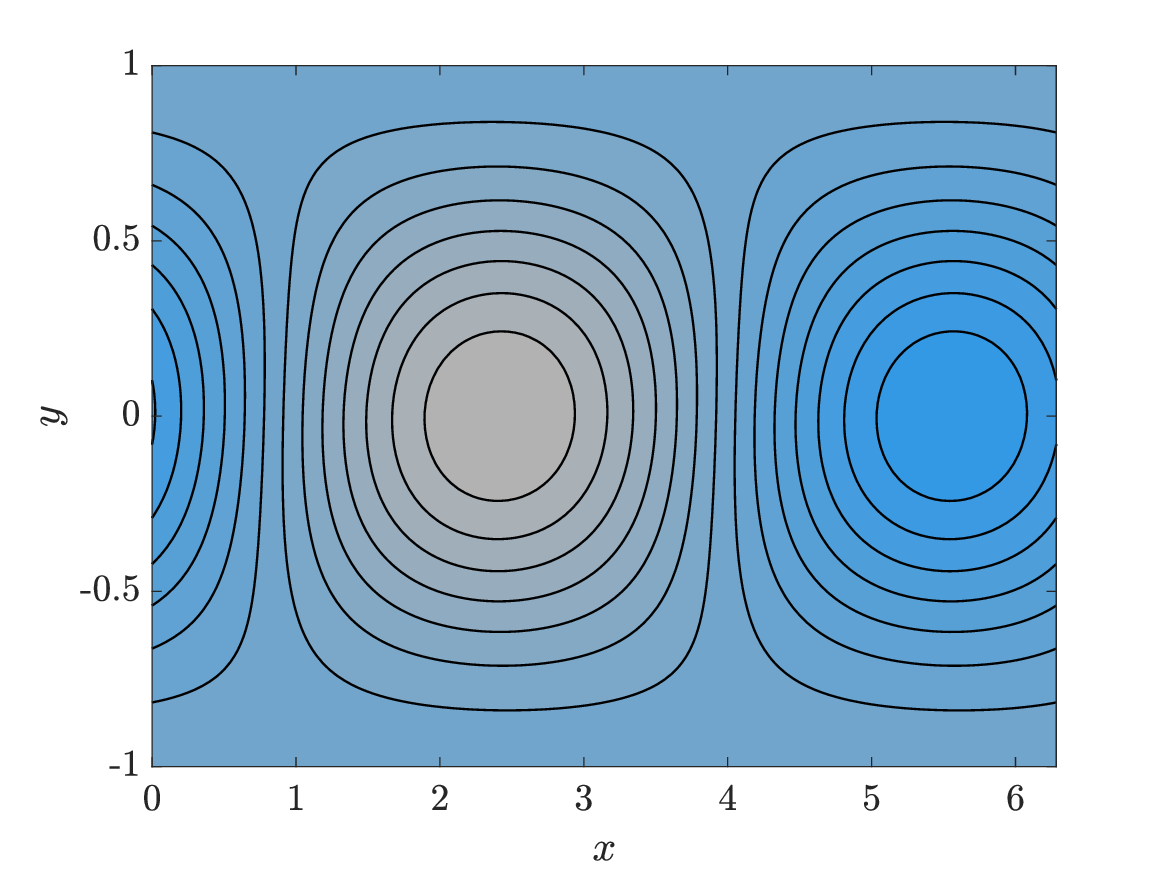}
        \caption{Streamlines for $k=1$.}
        \label{fig:13}
    \end{subfigure}
    \hfill
    \begin{subfigure}{0.49\linewidth}
        \centering
        \includegraphics[width=\linewidth,trim={0.65cm 0.2cm 0.65cm 0.2cm},clip]{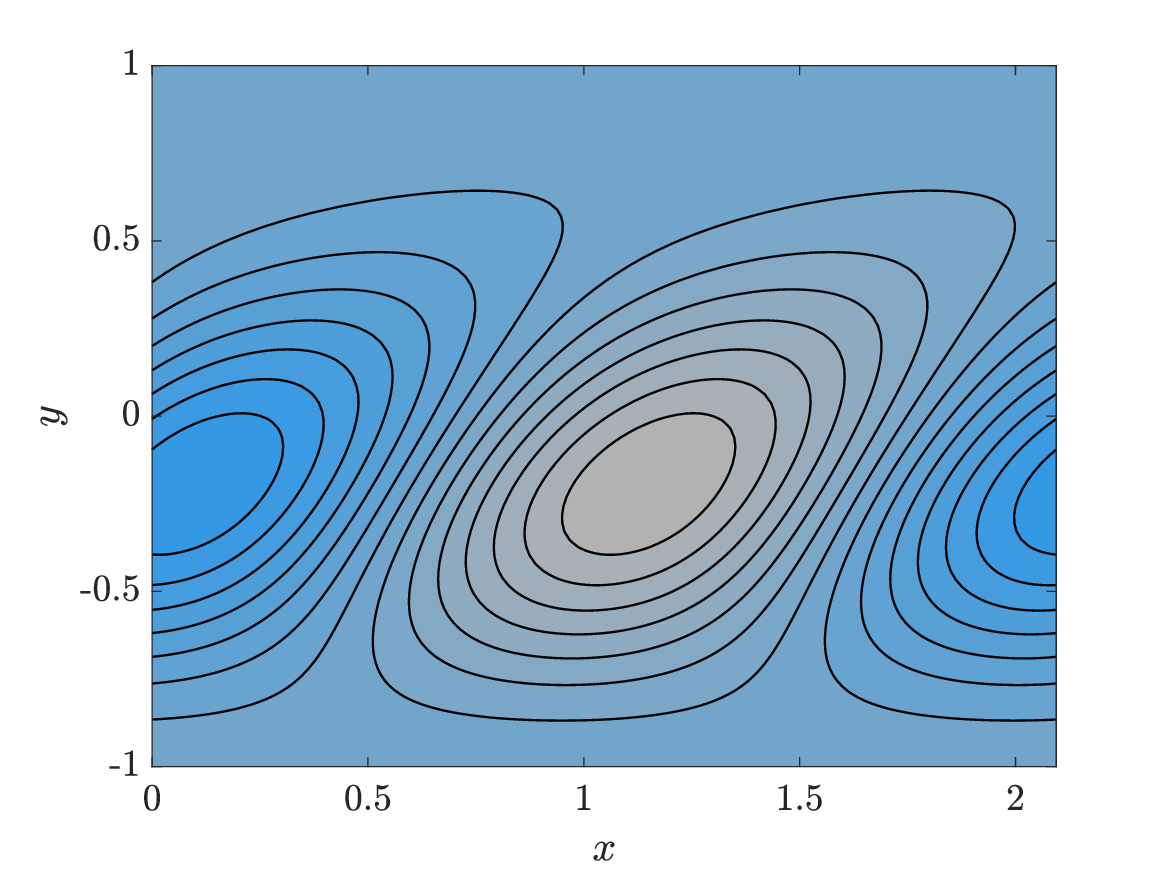}
        \caption{Streamlines for $k=6$.}
        \label{fig:14}
    \end{subfigure}

    \caption{Streamlines $\mathfrak{R}(\tilde{\psi}) = \text{const}$ for two choices of the wavenumber $k$ 
    ($Re = 15$, $Re_l = 0$, $a = 0.032$, $b = 1$, $\Gamma = 10^{-3}$, $n = -1.2$).}
    \label{fig:streamlines}
\end{figure}

To conclude our analysis of the case with prescribed stress on the top wall we plot the streamlines $\mathfrak{R}(\tilde{\psi})=const$ of the perturbation defined in \eqref{stream}  $k=1, 6$, $Re=15$, $Re_l=0$, $n=-1.2$, $a=0.032$, $b=1$, $\Gamma=10^{-3}$. From Figs. \ref{fig:13}, \ref{fig:14}, we see that the streamlines are qualitatively similar to those of Figs. \ref{fig:04quater}, \ref{fig:04five}.

\section{Conclusion}
In this work we derived the generalized stress power law model from the thermodynamic framework of Rajagopal and Srinivasa by constructing a rate of dissipation potential that is non-convex for a range of parameters. This approach provides a natural thermodynamic origin for constitutive relations that exhibit non-monotonic stress–strain rate behavior and may admit multiple steady state solutions. We analyzed the linearized stability of plane Couette flow for this class of fluids under both velocity and traction boundary conditions.

For velocity boundary conditions, the constitutive prescription may admit from one to three base states depending on the value of the relative Reynolds number. When three solutions exist, the two lying on ascending portions of the constitutive curve are unconditionally stable, whereas the solution on the descending portion is unconditionally unstable. Among the two stable branches, perturbations decay more rapidly on the third branch, which may be attributed to its viscosity being higher than that of the first branch.

For traction boundary conditions the base state is unique. However, if the solution lies on the descending branch the flow is unconditionally unstable, consistent with the behavior observed under velocity-driven flow, while solutions on the ascending branches remain unconditionally stable. One might expect that, at the inflection points where the base state switches from the first to the second ascending branch, a marginally stable state would arise. However, no neutrally stable solutions were observed. A key distinction between the two boundary conditions is that, when traction is prescribed, the stability depends solely on the Reynolds number $Re_u$ associated with the boundary on which traction is applied, and not on the relative Reynolds number. However, the Reynolds number associated with the boundary where the velocity is imposed $Re_l$ affects the decaying rate of the perturbation.

The analysis presented here is not intended to be exhaustive and has been restricted to the S-type non-monotonic case. We have focused on the instabilities arising solely from the constitutive response in the simplest shear flow. A natural continuation is to study the stability of solutions in Taylor–Couette flow, which is of greater experimental relevance and is expected to exhibit a much richer stability behavior due to the interplay between curvature and constitutive nonlinearities.

\section*{Acknowledgements}
The present work has been performed under the auspices of the Italian National Group for Mathematical Physics (GNFM-Indam). This work has been done under the framework PRIN 2022 project ``Mathematical modelling of heterogeneous systems''  financed by the European Union - Next Generation EU, CUP  B53D23009360006/B53D23009370006,   Project Code 2022MKB7MM, PNRR M4.C2.1.1.
\\
\bibliographystyle{elsarticle-num}
\bibliography{references}
\end{document}